\documentclass[aps,prd,preprint,superscriptaddress,floatfix]{revtex4-1}

\usepackage{xr}
\usepackage{zref}
\usepackage[T1]{fontenc}
\usepackage{amsmath}
\usepackage{amssymb}
\usepackage{units}
\usepackage{multirow}
\usepackage{graphicx}
\usepackage[sort&compress]{natbib}
\usepackage{geometry}
\usepackage{hyperref}
\setcitestyle{square,comma}
\usepackage{makecell}
\usepackage{bbm}

\newcommand{\MSbar}{\overline{\mathrm{MS}}}

\newcommand{\lat}{\mathrm{lat}}

\graphicspath{{figures/}}

\newcommand{\cu}{Physics Department, Columbia University, New York, 
	NY 10027, USA}
\newcommand{\soton}{School of Physics and Astronomy, University of Southampton, 
	Southampton SO17 1BJ, UK}
\newcommand{\edin}{SUPA, School of Physics, University of Edinburgh, Edinburgh EH9 3JZ, UK}
\newcommand{\xua}{School of Physics, Peking University, Beijing 100871, China}
\newcommand{\xub}{Center for High Energy Physics, Peking University, Beijing 100871, China}
\newcommand{\xuc}{Collaborative Innovation Center of Quantum Matter, Beijing 100871, China}

\begin{document}

\title{First exploratory calculation of the long-distance contributions to the rare kaon decays $K\rightarrow\pi\ell^{+}\ell^{-}$}

\author{Norman~H.~Christ}\affiliation{\cu}
\author{Xu~Feng}\affiliation{\cu}\affiliation{\xua}\affiliation{\xub}\affiliation{\xuc}
\author{Andreas~J\"{u}ttner}\affiliation{\soton}
\author{Andrew~Lawson}\affiliation{\soton}
\author{Antonin~Portelli}\affiliation{\soton}\affiliation{\edin}
\author{Christopher~T.~Sachrajda}\affiliation{\soton}
\collaboration{RBC and UKQCD collaborations}
\date{\today}
\pacs{PACS}
    
\begin{abstract}
	The rare decays of a kaon into a pion and a charged lepton/antilepton pair proceed via a flavor changing neutral current and therefore may only be induced beyond tree level in the Standard Model. This natural suppression makes these decays sensitive to the effects of potential new physics. The $CP$-conserving $K\to\pi \ell^+\ell^-$ decay channels however are dominated by a single-photon exchange; this involves a sizeable long-distance hadronic contribution which represents the current major source of theoretical uncertainty. Here we outline our methodology for the computation of the long-distance contributions to these rare decay amplitudes using lattice QCD and present the numerical results of the first exploratory studies of these decays in which all but the disconnected diagrams are evaluated. The domain wall fermion ensembles of the RBC and UKQCD Collaborations are used, with a pion mass of $M_{\pi}\sim\unit[430]{MeV}$ and a kaon mass of $M_{K}\sim\unit[625]{MeV}$. In particular we determine the form factor, $V(z)$, of the $K^+\to\pi^+\ell^+\ell^-$ decay from the lattice at small values of $z=q^2/M_{K}^{2}$, obtaining $V(z)=1.37(36),\, 0.68(39),\, 0.96(64)$ for the three values of $z=-0.5594(12),\, -1.0530(34),\, -1.4653(82)$ respectively.
\end{abstract}

\maketitle

\newpage

\tableofcontents

\section{Introduction}

The rare kaon decays $K\rightarrow \pi \ell^+ \ell^-$ and $K\rightarrow \pi \nu \bar{\nu}$ are flavor changing neutral current processes, which are naturally suppressed in the Standard Model as they first arise only as second-order electroweak processes. This suppression makes them ideal probes for new physics effects. 

One significant difficulty in the theoretical understanding of second-order weak processes is that there may be significant contributions when the two electroweak vertices are separated by distances as large as $1/\Lambda_{\mathrm{QCD}}$. These long-distance effects contain nonperturbative contributions, hence a complete theoretical study of these processes can be achieved only by utilizing nonperturbative methods such as lattice QCD. However $K\to\pi\nu\bar\nu$ decays are short-distance dominated, as the absence of photon exchange diagrams suppresses the long-distance contributions. These processes feature a quadratic (hard) GIM (Glashow-Iliopoulos-Maiani) mechanism~\cite{PhysRevD.2.1285}, such that the loop diagrams that mediate the decays depend quadratically on the mass of the quark entering the loop. This plays a part in enhancing the short-distance contribution involving heavy quarks. Furthermore for the direct $CP$-violating component of the decay $K_L\to\pi^0\nu\bar{\nu}$, the amplitude is dependent upon the Cabibbo-Kobayashi-Maskawa (CKM) matrix factor $\textrm{Im}\left(\lambda_q\right)$ (where $\lambda_q=V^{*}_{sq}V_{qd}$), which significantly suppresses the up and charm contributions. As a result, this decay is entirely dominated by loops involving the top quark.

The story for $K\to\pi \ell^+\ell^-$ processes is considerably different, as they may be mediated by a single-photon exchange, whose amplitude is determined by nonperturbative, long-distance physics. The $CP$-conserving processes $K_S\to\pi^0\ell^+\ell^-$ and $K^+\to\pi^+\ell^+\ell^-$ are dominated by the single-photon exchange amplitude, where the short-distance top quark contribution is suppressed by the CKM factor $\textrm{Re}\left(\lambda_t\right)$ and even a potentially large light-quark short-distance contribution is cut off at the charm quark Compton wave length by a logarithmic GIM cancellation. The Z-exchange and box-diagram amplitudes in these processes are suppressed by a factor of $1/M_{Z}^{2}$, and are comparatively negligible. Although the short-distance top quark contribution is enhanced by a factor of $m_{t}^{2}$ (which compensates for the $1/M_{Z}^{2}$ suppression), the CKM factor $\textrm{Re}\left(\lambda_t\right)$ nevertheless suppresses the top quark contribution. For the case of $K_L\to\pi^0\ell^+\ell^-$ the long-distance contributions to the component that directly violates $CP$ are again suppressed by CKM matrix factors. There is also however a significant long-distance contribution originating from indirect $CP$ violation, and a $CP$-conserving contribution from $K_L\to\pi^0\gamma^\ast\gamma^\ast$ with $\gamma^\ast\gamma^\ast\to \ell^+\ell^-$ rescattering.

Rare kaon decays have received much focus from experimentalists for many years. Traditionally the decay channels $K\rightarrow\pi \nu \bar{\nu}$ have been more of an interest owing to the short-distance dominance, and hence theoretical control of the hadronic effects, described above. The detection of such events has proven to be a significant experimental challenge. At present there exist dedicated experiments at J-PARC (KOTO)~\cite{Yamanaka:2012yma} and CERN (NA62)~\cite{Anelli:2005ju} which primarily aim to measure the $K_L\rightarrow\pi^0 \nu \bar{\nu}$ and $K^+\rightarrow\pi^+ \nu \bar{\nu}$ branching ratios respectively to within $10\%$. Although long-distance contributions are expected to account for a small percentage of the overall amplitude for $K^+\rightarrow\pi^+ \nu \bar{\nu}$ decays, a lattice QCD calculation may play an important role in rigorously controlling the size of this theoretical uncertainty. The prospects for such a lattice calculation have been discussed recently in~\cite{Christ:2016eae}.

On the other hand, branching ratios for $K^+\rightarrow \pi^+ \ell^+ \ell^-$ processes are known to a considerably higher degree of accuracy: $\mathrm{Br}\left(K^{\pm}\rightarrow\pi^{\pm}e^+e^-\right)=3.14(10)\times10^{-7}$~\cite{Batley:2009aa} and $\mathrm{Br}\left(K^{\pm}\rightarrow\pi^{\pm}\mu^+\mu^-\right)=9.62(25)\times10^{-8}$~\cite{Batley:2011zz}. It is likely that the NA62 experiment will also determine these branching ratios to a higher precision. With higher statistics there is hope that the experiment may be sensitive to lepton flavor universality violation in rare kaon decays~\cite{Crivellin:2016vjc}. $K_S\rightarrow \pi^0 \ell^+ \ell^-$ decays however are more challenging to measure, although their detection is important for calculating the indirect $CP$-violating contribution to $K_L\rightarrow \pi^0 \ell^+ \ell^-$ decays via the chain $K_L\rightarrow K_1\rightarrow \pi^0 \ell^+ \ell^-$, where $K_1$ is the $CP$-even component of $K_L$. The branching ratios are currently only known with $\sim50\%$ errors: $\mathrm{Br}\left(K_S\rightarrow\pi^{0}e^+e^-\right)=\left(5.8^{+2.9}_{-2.4}\right)\times10^{-9}$~\cite{Batley:2003mu} and $\mathrm{Br}\left(K_S\rightarrow\pi^{0}\mu^+\mu^-\right)=\left(2.9^{+1.5}_{-1.2}\right)\times10^{-9}$~\cite{Batley:2004wg}. Given the difficulty of the experimental measurement, there exists a good opportunity to extract this result instead from lattice QCD simulations. In addition, such a lattice calculation will determine the phase of the indirect $CP$-violating amplitude, which cannot be determined from an experimental measurement of the $K_S\to\pi^0\ell^+\ell^-$ branching ratio.

On the lattice we aim to compute the dominant long-distance contribution to the matrix element  $K\rightarrow\pi\gamma^*$ (i.e. the single-photon exchange channel). The plans for such a calculation have been discussed in a recent paper~\cite{Christ:2015aha}, building on the work of~\cite{Isidori:2005tv}. Our primary focus is the $K^+\to\pi^+\gamma^*\to\pi^+\ell^+\ell^-$ decay, although we will also comment briefly on the decay with neutral hadrons. Previous theoretical work on this decay is mainly based on chiral perturbation theory (ChPT) and has led to various parametrizations of the form factor for the decay; the status of this work has been reviewed in~\cite{Cirigliano:2011ny}. Coefficients in these parametrizations have been obtained from fits to experimental data~\cite{Batley:2003mu,Batley:2004wg,Batley:2009aa,Batley:2011zz}. An early opportunity for lattice QCD is to use our simulation data to test the reliability of this previous theoretical work.

The calculation we present in this paper is the first exploratory attempt at a nonperturbative lattice QCD calculation of $K\rightarrow \pi \ell^+ \ell^-$ amplitudes. The possibility of such a calculation was first introduced in~\cite{Isidori:2005tv}, where it was shown that lattice methods can in principle be used to compute such decay amplitudes. These ideas were developed further in~\cite{Christ:2015aha}, where the details of the analysis to extract $K\rightarrow \pi \ell^+ \ell^-$ matrix elements using renormalized operators were introduced, with full control of ultraviolet divergences. This necessitates the introduction of a charm quark in the calculation, such that logarithmic divergences cancel by the GIM mechanism.  Our objective is to demonstrate how the results of~\cite{Isidori:2005tv,Christ:2015aha} can be applied in actual numerical simulations to extract the desired physical information. In this paper we report on the results of our exploratory numerical simulations of the rare kaon decay $K^+\rightarrow \pi^+ \ell^+ \ell^-$ using the domain wall fermion (DWF) ensembles of the RBC and UKQCD Collaborations~\cite{Aoki:2010dy}.

The layout of this paper is as follows. In Sec.~\ref{sec:Ops} we outline the lattice operators necessary to study $K\to\pi\ell^+\ell^-$ decays, briefly summarizing the work of~\cite{Isidori:2005tv}. In Sec.~\ref{sec:Analysis} we follow and build on~\cite{Christ:2015aha} to give a detailed discussion of the analysis methods necessary to extract the rare kaon decay amplitudes from the lattice results. In Sec.~\ref{sec:Simulation} we give details of the implementation of the lattice simulation we performed to obtain our numerical results. These numerical results are discussed in Sec.~\ref{sec:Results}. In Sec.~\ref{sec:Form_Factor} we briefly summarize existing theoretical results for $K^+\rightarrow \pi^+ \ell^+ \ell^-$ decays, before making use of our lattice results to outline how we can test existing $\mathcal{O}\left(p^4\right)$ ChPT and experimental results, once all systematic effects in our calculation are controlled. Finally in Sec.~\ref{sec:Conclusion} we present our conclusions. We remark that all dimensionful quantities appearing in this paper are expressed in lattice units unless otherwise stated.

\section{\label{sec:Ops}Operators and Contractions}

The expression for the long-distance Minkowski amplitude we wish to compute is given by
\begin{equation}
\mathcal{A}^{i}_{\mu}\left(q^{2}\right)=\int d^{4}x\left\langle \pi^{i}\left(\mathbf{p}\right)|T\left[J_{\mu}\left(0\right)\mathcal{H}_{W}\left(x\right)\right]|K^{i}\left(\mathbf{k}\right)\right\rangle ,\label{eq:Minkowski_amplitude}
\end{equation}
where $q=k-p$ and $i=+,0$. Using electromagnetic gauge invariance this nonlocal
matrix element can be written as
\begin{equation}
\mathcal{A}^{i}_{\mu}\left(q^{2}\right)\equiv -i\, G_F\dfrac{V^i\left(z\right)}{\left(4\pi\right)^{2}}\left(q^{2}\left(k+p\right)_{\mu}-\left(M_{K}^{2}-M_{\pi}^{2}\right)q_{\mu}\right),\label{eq:mat_elem_form_fac}
\end{equation}
where nonperturbative QCD effects are contained in the form factor
$V^i\left(z\right)$, $z=q^{2}/M_{K}^{2}$ (note we are using the notation of Ref.~\cite{Cirigliano:2011ny} for $V^i\left(z\right)$).

The four-flavor effective weak Hamiltonian relevant to the transition $s\rightarrow d\ell^{+}\ell^{-}$
renormalized at a scale $\mu$ with $M_{W}\gg\mu>m_{c}$ is defined by~\cite{Buchalla:1995vs}
\begin{equation}
\mathcal{H}_{W}=\dfrac{G_{F}}{\sqrt{2}}V_{us}^{*}V_{ud}\left(\sum_{j=1}^{2}C_{j}\left(Q_{j}^{u}-Q_{j}^{c}\right)+\sum_{j=3}^{8}C_{j}Q_{j}+\mathcal{O}\left(\dfrac{V_{ts}^{*}V_{td}}{V_{us}^{*}V_{ud}}\right)\right).\label{eq:weak_hamiltonian}
\end{equation}
In practice the operators $Q_{3,\dots,8}$ may be neglected as the corresponding Wilson coefficients $C_{3,\dots,8}$ are much smaller than those of $Q_1$ and $Q_2$~\cite{Buchalla:1995vs,Isidori:2005tv}. We will therefore consider only these two operators defined as
\begin{eqnarray}
Q_{1}^{q}=\left(\bar{s}_{i}\gamma_{\mu}^{L}d_{i}\right)\left(\bar{q}_{j}\gamma^{L,\mu}q_{j}\right), & \quad & Q_{2}^{q}=\left(\bar{s}_{i}\gamma_{\mu}^{L}d_{j}\right)\left(\bar{q}_{j}\gamma^{L,\mu}q_{i}\right),
\end{eqnarray}
where $i,j$ are summed color indices and $\gamma_{\mu}^{L}=\gamma_{\mu}\left(1-\gamma_{5}\right)$. For clarity, in later sections we will refer to the operator
\begin{align}
	H_{W}=\sum_{j=1}^{2}C_{j}\left(Q_{j}^{u}-Q_{j}^{c}\right),\label{eq:weak_operator}
\end{align}
and the prefactor $G_{F}V_{us}^{*}V_{ud}/\sqrt{2}$ will be inserted later. In the lattice computations we start by determining the matrix elements of these bare lattice operators and then use nonperturbative renormalization to obtain them in the RI-SMOM scheme. We subsequently use perturbation theory to match with the Wilson coefficients for the $\MSbar$ scheme, which are known at next-to-leading order~\cite{Buchalla:1995vs}. The matching formulas for this step have been previously calculated and presented in Ref.~\cite{Lehner:2011fz}. The procedure used for this calculation is identical to that used for the renormalization of the $H_W$ operator in the calculation of the $K_{L}-K_{S}$ mass difference, as discussed in Sec. VII of Ref.~\cite{Christ:2012se}.

The electromagnetic current $J_{\mu}$ in Eq. (\ref{eq:Minkowski_amplitude}) is the standard flavor-diagonal operator
\begin{equation}
J_{\mu}=\dfrac{1}{3}\left(2V_{\mu}^{u}-V_{\mu}^{d}-V_{\mu}^{s}+2V_{\mu}^{c}\right),
\end{equation}
where $V_{\mu}^{q}$ is the conserved lattice vector current for the flavor $q$. For our choice of action we use the Shamir domain wall conserved current~\cite{Shamir:1993yf}.

\subsection{Wick contractions}

\begin{figure}
	\begin{centering}
		\begin{tabular}{cc}
			\includegraphics[scale=0.9]{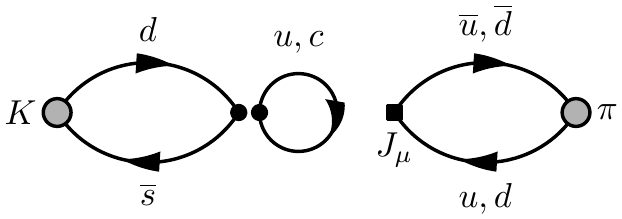} & \includegraphics[scale=0.9]{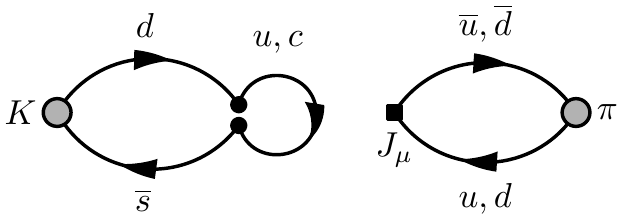}\tabularnewline
		\end{tabular}
		\par\end{centering}
	
	\protect\caption{\label{fig:pi0_contractions}The additional two classes of diagrams obtained after performing the Wick contractions of the neutral pion and kaon interpolating operators with the $H_{W}$ operator.}
\end{figure}

Inserting the weak Hamiltonian Eq. (\ref{eq:weak_operator})
and the electromagnetic current into Eq. (\ref{eq:Minkowski_amplitude}),
we can perform all Wick contractions to produce the 20 diagrams that must be computed. It is convenient to start by performing the Wick contractions for the insertion of only the operator $H_{W}$ to obtain the four
different classes of diagrams shown in Fig.~\ref{fig:H_W_contractions}. 
Within each class there are then five possible diagrams, obtained by inserting the electromagnetic current in all possible ways. First the current can be inserted on any of the quark propagators in each class. There is also the possibility of the self-contraction of the current to produce a disconnected diagram, which corresponds to a photon being emitted from a sea quark loop. We illustrate the five insertions for the $C$ class in Fig.~\ref{fig:J_contractions}. We remark that for the neutral case $K_{S}\rightarrow\pi^{0}\ell^+\ell^-$ we can also contract the two quarks within the pion to produce two disconnected diagram topologies shown in Fig.~\ref{fig:pi0_contractions}. A full list of diagrams can be found in Ref.~\cite{Christ:2015aha}.

When the current is inserted in the loop of the $S$ and $E$ diagrams, there appear to be quadratically divergent contributions as the operators $J_{\mu}$ and $H_{W}$ approach each other \cite{Isidori:2005tv,Christ:2015aha}. As we simulate with a conserved current we can rely on electromagnetic gauge invariance to reduce the degree of divergence by two dimensions (owing to a transversality factor of $q^2 g_{\mu\nu}-q_\mu q_\nu$), leaving at most a logarithmic divergence. This remaining divergence is canceled by introducing a charm quark (as displayed in Fig.~\ref{fig:H_W_contractions}) and exploiting the GIM mechanism~\cite{PhysRevD.2.1285}. We remark that the inclusion of the charm is not merely for convenience: it is necessary to perform the lattice calculation with four flavors to be confident in the accuracy of the final result. The contribution of the charm quark to this decay can be roughly estimated using the formulas of Ref.~\cite{Inami:1980fz}. Such an estimate suggests that the new diagrams obtained by introducing an electromagnetic current vertex into the charm and up loops in the $S$ and $E$ graphs of Fig.~\ref{fig:H_W_contractions} may give a relatively large effect. Such an effect is best determined by a complete lattice calculation of such GIM-subtracted contributions, which necessarily contains a valence charm quark.

\begin{center}
	\begin{figure}
		\begin{centering}
			\begin{tabular}{cccc}
				\includegraphics[scale=0.6]{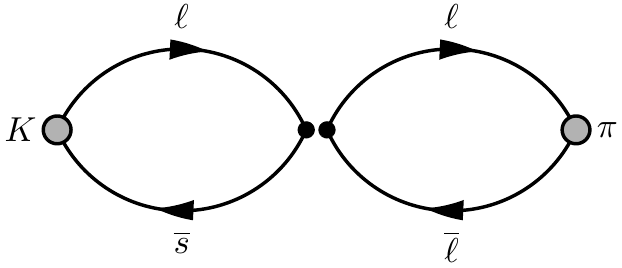} & \includegraphics[scale=0.6]{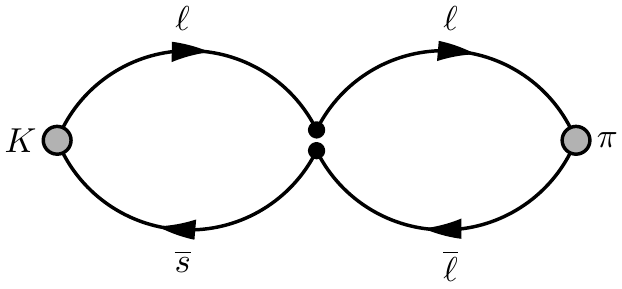} & \includegraphics[scale=0.6]{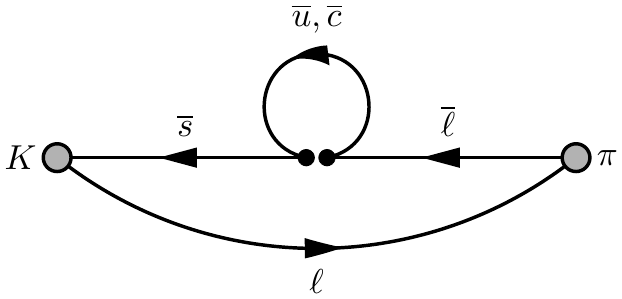} & \includegraphics[scale=0.6]{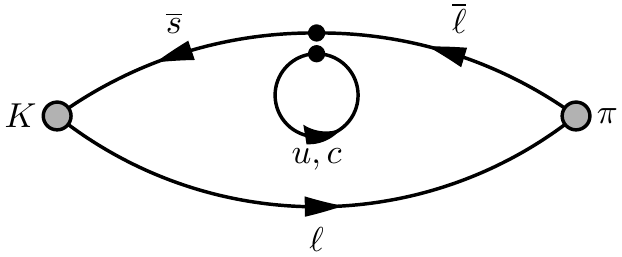}\tabularnewline
				$W$ & $C$ & $S$ & $E$\tabularnewline
				(Wing) & (Connected) & (Saucer) & (Eye)\tabularnewline
			\end{tabular}
			\par\end{centering}
		
		\protect\caption{\label{fig:H_W_contractions}The four classes of diagrams obtained after performing the Wick contractions of the charged pion and kaon interpolating operators with the $H_{W}$ operator.}
	\end{figure}
	
	\begin{figure}
		\begin{centering}
			\begin{tabular}{ccc}
				\includegraphics[scale=0.6]{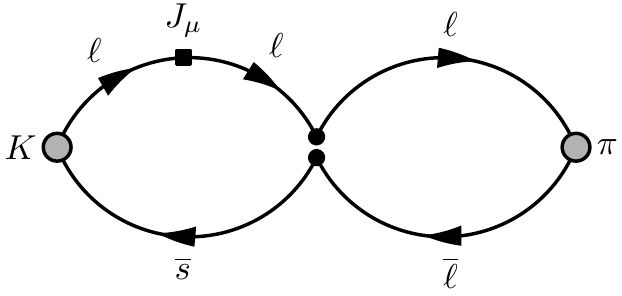} & \includegraphics[scale=0.6]{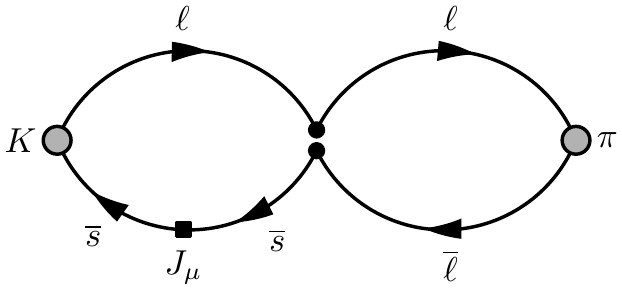} & \includegraphics[scale=0.6]{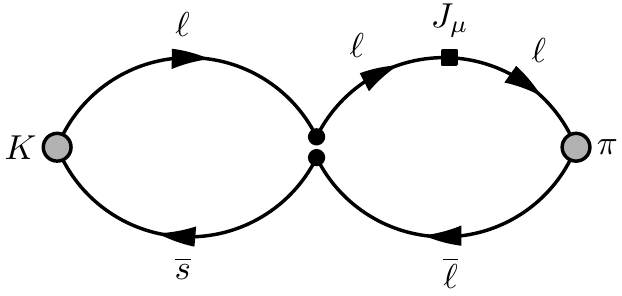}\tabularnewline
				\multicolumn{3}{c}{\includegraphics[scale=0.6]{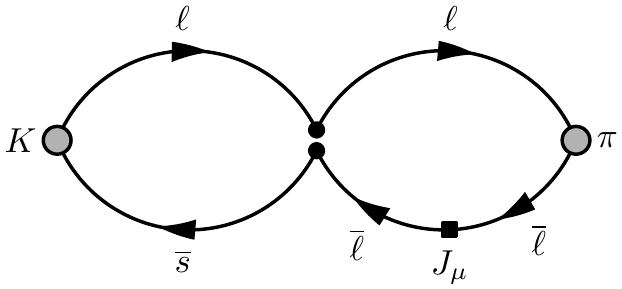}\includegraphics[scale=0.6]{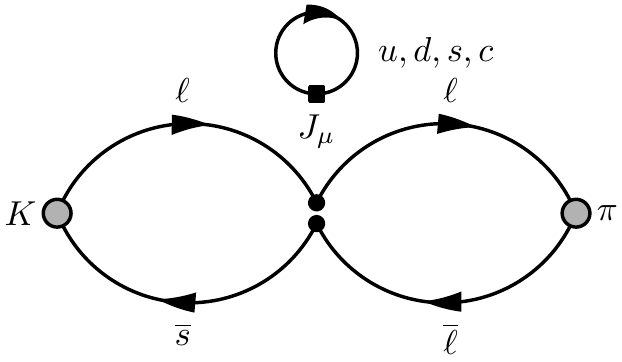}}\tabularnewline
			\end{tabular}
			\par\end{centering}
		
		\caption{\label{fig:J_contractions}The five possible current insertions for the $C$ class of diagrams.}
	\end{figure}
	
	\par\end{center}

\section{\label{sec:Analysis}Determination of the Matrix Element}

In this section we outline the analysis techniques necessary to extract rare kaon decay amplitudes from the four-point (4pt) correlators measured in our lattice simulation. We begin by discussing the extraction of Euclidean amplitudes in the continuum, followed by a discussion of the additional considerations we must make in discrete spacetime.

\subsection{Continuum Euclidean correlators}

In order to measure the amplitude given by Eq. (\ref{eq:Minkowski_amplitude}) on the lattice, we start by defining the "unintegrated" 4pt correlator
\begin{equation}
\Gamma_{\mu}^{\left(4\right)}\left(t_{H},t_{J},\mathbf{k},\mathbf{p}\right)=\int d^3\mathbf{x}\int d^3\mathbf{y}\;e^{-i\mathbf{q}\cdot\mathbf{x}}\left\langle \phi_{\pi}\left(t_{\pi},\mathbf{p}\right)T\left[J_{\mu}\left(t_{J},\mathbf{x}\right)H_{W}\left(t_{H},\mathbf{y}\right)\right]\phi_{K}^{\dagger}\left(t_K,\mathbf{k}\right)\right\rangle ,\label{eq:4pt_corr}
\end{equation}
where the operator $\phi_{P}\left(t,\mathbf{p}\right)$ is the annihilation operator for a pseudoscalar meson $P$ with momentum $\mathbf{p}$ at a time $t$. To obtain the decay amplitude we must consider the integrated 4pt correlator,
\begin{equation}
I_{\mu}\left(T_{a},T_{b},\mathbf{k},\mathbf{p}\right)=e^{-\left(E_{\pi}\left(\mathbf{p}\right)-E_{K}\left(\mathbf{k}\right)\right)t_{J}}\int_{t_{J}-T_{a}}^{t_{J}+T_{b}}dt_H\;\tilde{\Gamma}_{\mu}^{\left(4\right)}\left(t_{H},t_{J},\mathbf{k},\mathbf{p}\right),\label{eq:integrated_corr}
\end{equation}
in the limit $T_{a},T_{b}\rightarrow\infty$~\cite{Christ:2015aha}. We define $\tilde{\Gamma}_{\mu}^{\left(4\right)}$ as the "reduced" correlator after dividing out the source/sink factors and normalizations which do not contribute to the final amplitude, i.e.
\begin{equation}\label{eq:Gamma_4_tilde}
\tilde{\Gamma}_{\mu}^{\left(4\right)}=\dfrac{\Gamma_{\mu}^{\left(4\right)}}{Z_{\pi K}},\quad Z_{\pi K}=\dfrac{Z_{\pi}Z_{K}^{\dagger}L^3}{4E_{\pi}\left(\mathbf{p}\right)E_{K}\left(\mathbf{k}\right)}e^{-t_{\pi}E_{\pi}\left(\mathbf{p}\right)+t_{K}E_{K}\left(\mathbf{k}\right)},
\end{equation}
with $Z_{\pi}=\big\langle \pi\left(\mathbf{p}\right)|\phi_{\pi}\left(\mathbf{p}\right)|0\big\rangle$, $Z^{\dagger}_{K}=\big\langle 0|\phi_{K}^{\dagger}\left(\mathbf{k}\right)|K\left(\mathbf{k}\right)\big\rangle$, and $E_K\left(\mathbf{k}\right)$ and $E_{\pi}\left(\mathbf{p}\right)$ are the initial state kaon and final state pion energies respectively. These parameters can be extracted from fits of the relevant two-point (2pt) correlation functions. We account for the factor of $L^3$ (i.e. the spatial volume) as we integrate both $\mathbf{x}$ and $\mathbf{y}$ over all space. The exponential factor outside the integral in Eq. (\ref{eq:integrated_corr}) effectively translates the decay to $t_J=0$ (as is allowed by translational invariance); we will therefore omit further $t_J$ dependence from our expressions.

The spectral decomposition of the unintegrated 4pt correlator for $t_K\ll t_H$ and $t_H\ll t_{\pi}$ can be written as:
\begin{align}
\tilde{\Gamma}_{\mu}^{\left(4\right)}\left(t_{H},\mathbf{k},\mathbf{p}\right) = & \begin{cases} \int_{0}^{\infty} dE\;\dfrac{\rho\left(E\right)}{2E}\left\langle \pi\left(\mathbf{p}\right)|J_{\mu}|E,\mathbf{k}\right\rangle \left\langle E,\mathbf{k}|H_{W}|K\left(\mathbf{k}\right)\right\rangle e^{-\left(E_K(\mathbf{k})-E\right)t_H}, & t_H < 0, \\[0.5ex]
\int_{0}^{\infty} dE\;\dfrac{\rho_{S}\left(E\right)}{2E}\left\langle \pi\left(\mathbf{p}\right)|H_{W}|E,\mathbf{p}\right\rangle \left\langle E,\mathbf{p}|J_{\mu}|K\left(\mathbf{k}\right)\right\rangle e^{-\left(E-E_{\pi}(\mathbf{p})\right)t_H}, & t_H > 0, \end{cases}\label{eq:unint_spec_rep}
\end{align}
where the functions $\rho\left(E\right)$ and $\rho_{S}\left(E\right)$ are the relevant spectral densities which select states with strangeness $S=0$ and $S=1$ respectively. The integral over $t_H$ in Eq. (\ref{eq:integrated_corr}) can thus be computed analytically to obtain
\begin{align}
I_{\mu}\left(T_{a},T_{b},\mathbf{k},\mathbf{p}\right) = & -\int_{0}^{\infty} dE\;\dfrac{\rho\left(E\right)}{2E}\dfrac{\left\langle \pi\left(\mathbf{p}\right)|J_{\mu}|E,\mathbf{k}\right\rangle \left\langle E,\mathbf{k}|H_{W}|K\left(\mathbf{k}\right)\right\rangle }{E_{K}\left(\mathbf{k}\right)-E}\left(1-e^{\left(E_{K}\left(\mathbf{k}\right)-E\right)T_{a}}\right)\nonumber \\
& +\int_{0}^{\infty} dE\;\dfrac{\rho_{S}\left(E\right)}{2E}\dfrac{\left\langle \pi\left(\mathbf{p}\right)|H_{W}|E,\mathbf{p}\right\rangle \left\langle E,\mathbf{p}|J_{\mu}|K\left(\mathbf{k}\right)\right\rangle }{E-E_{\pi}\left(\mathbf{p}\right)}\left(1-e^{-\left(E-E_{\pi}\left(\mathbf{p}\right)\right)T_{b}}\right).\label{eq:int_spec_rep_cont}
\end{align}
The rare kaon decay amplitude we wish to calculate corresponds to the constant terms in the above equation (i.e. those that do not depend on the exponentials in $T_a$ and $T_b$)~\cite{Christ:2015aha}. The states $\left|E,\mathbf{p}\right\rangle$ in the second line of Eq. (\ref{eq:int_spec_rep_cont}) must have the flavor quantum numbers of a kaon, i.e. $S=1$, and thus all possible states will have $E>E_{\pi}\left(\mathbf{p}\right)$;
given a sufficiently large $T_{b}$ this half of the integral should
converge to the appropriate value. However the states $\left|E,\mathbf{k}\right\rangle$ in the first line have the
quantum numbers of a pion. For physical pion and kaon masses there
are three permitted intermediate states with $E<E_K\left(\mathbf{k}\right)$ (namely one, two and three pion states), which will cause the integral
to diverge with increasing $T_{a}$. These exponentially growing contributions from these three types of intermediate states do not contribute to the overall decay width and therefore must be removed in order to extract the relevant Minkowski amplitude,
\begin{equation}
\mathcal{A}_{\mu}\left(q^{2}\right)=-i\dfrac{G_{F}}{\sqrt{2}}V_{us}^{*}V_{ud}\lim_{T_{a},T_{b}\rightarrow\infty}\tilde{I}_{\mu}\left(T_{a},T_{b},\mathbf{k},\mathbf{p}\right),\label{eq:Euclid_to_Minkowski}
\end{equation}
where $\tilde{I}_{\mu}$ indicates the integrated 4pt correlator after
subtracting the exponentially growing contributions~\cite{Christ:2015aha}.

\subsection{Lattice implementation}

In our lattice simulation we compute the correlator in Eq. (\ref{eq:4pt_corr}) in a finite volume at a finite lattice spacing; for the purposes of our analysis it is useful to translate these continuum, infinite-volume formulas into their discrete, finite-volume counterparts. To make the difference between the two clear, we will not suppress factors of the lattice spacing for the remainder of this section.

The spectral density $\rho\left(E\right)$ in finite volume can be expressed as $\rho\left(E\right)=\sum_{n}2E_n\delta\left(E-E_n\right)$ [and similarly for $\rho_{S}\left(E\right)$]; our phase space integral is hence reduced to a sum over a finite number of states labeled by $n$. The spatial integrals in Eq. (\ref{eq:4pt_corr}) are replaced by sums over the spatial extent of the lattice. Similarly the integral in Eq. (\ref{eq:integrated_corr}) can be replaced by a sum. 

The replacement of integrals over $t_H$ by sums in our lattice calculation corresponds to the replacement
\begin{align}
\int_{-T_a}^{0}dt_H\to  a\sum_{t_H=-T_a}^{0},&\quad
\int_{0}^{T_b}dt_H\to  a\sum_{t_H=0}^{T_b}.
\end{align}
The sum runs over increments of the lattice spacing, $a$. We remark that the point at $t_H=t_J=0$ should not be double counted when the two halves of the integral are added together; this is intrinsically related to how the time ordering operator is implemented on the lattice. Because the operators $H_W$ and $J_{\mu}$ commute at $t_H=0$, a proper treatment is to average the two choices of time ordering at this point. In the following analysis the point at $t_H=0$ is thus weighted by a half; when the two sums are added together the correct result is obtained.

We now introduce the compact notation
\begin{align}
\Delta_{n}^{a}=  E_K\left(\mathbf{k}\right) - E_n,&\quad
\Delta_{m}^{b}=  E_m - E_{\pi}\left(\mathbf{p}\right),
\end{align}
where $n$ and $m$ label the finite volume states contained in the finite volume spectral densities $\rho\left(E\right)$ and $\rho_{S}\left(E\right)$ respectively, and $a$ and $b$ label which time ordering of the 4pt function the state appears for. The relevant sums corresponding to the integral of Eq.~(\ref{eq:4pt_corr}) can be evaluated as a geometric series, i.e.
\begin{align}
a\sum_{t_H=-T_a}^{0}e^{-\Delta_{n}^{a} t_H}=a\dfrac{1+e^{a\Delta_{n}^{a\vphantom{b}}}\left(1-2e^{\Delta_{n}^{a\vphantom{b}}T_{a\vphantom{b}}}\right)}{2\left(1-e^{a\Delta_{n}^{a\vphantom{b}}}\right)}, &\quad
a\sum_{t_H=0}^{T_b}e^{-\Delta_{m}^{b} t_H}=a\dfrac{1+e^{-a\Delta_{m}^{b}}\left(1-2e^{-\Delta_{m}^{b}T_{b}}\right)}{2\left(1-e^{-a\Delta_{m}^{b}}\right)}. \label{eq:sum_1}
\end{align}
To understand the impact of this analysis, it is instructive to expand the terms in Eq. (\ref{eq:sum_1}) that depend on $T_a$ and $T_b$. Expanding in powers of the lattice spacing, the unphysical contributions take the form:
\begin{align}
	-a\dfrac{e^{a\Delta_{n}^{a}}}{1-e^{a\Delta_{n}^{a}}}e^{\Delta_{n}^{a}T_{a}}&=\left(1+\dfrac{a\Delta_{n}^{a}}{2}+\dfrac{\left(a\Delta_{n}^{a}\right)^{2}}{12}+\mathcal{O}\left(a^{3}\right)\right)\dfrac{e^{\Delta_{n}^{a}T_{a}}}{\Delta_{n}^{a}}, \label{eq:sum_1_exp} \\
	-a\dfrac{e^{-a\Delta_{m}^{b}}}{1-e^{-a\Delta_{m}^{b}}}e^{-\Delta_{m}^{b}T_{b}}&=\left(-1+\dfrac{a\Delta_{m}^{b}}{2}-\dfrac{\left(a\Delta_{m}^{b}\right)^{2}}{12}+\mathcal{O}\left(a^{3}\right)\right)\dfrac{e^{-\Delta_{m}^{b}T_{b}}}{\Delta_{m}^{b}}. \label{eq:sum_2_exp}
\end{align}
This analysis demonstrates the expectation that the sum reproduces the continuum expectation, up to discretization effects starting at $\mathcal{O}\left(a\right)$. Neglecting these effects would result in an incomplete removal of the exponentially growing behavior, which could introduce a significant systematic effect into our analysis and thus should be avoided. We stress however that the physical matrix element itself, i.e. the contribution of those terms in Eq.~(\ref{eq:int_spec_rep_cont}) without the factors of $e^{\Delta_m^aT_a}$ or $e^{\Delta_m^bT_b}$, is free of $\mathcal{O}\left(a\right)$ errors as is guaranteed by our prescription of domain wall fermions.

We can thus write the final expression for our integrated lattice correlator,
\begin{align}
I_{\mu}\left(T_{a},T_{b},\mathbf{k},\mathbf{p}\right) = &\: a\sum_{n}\dfrac{1}{2E_{n}}\dfrac{\mathcal{M}_{\mu}^{J,n\rightarrow \pi}\left(\mathbf{k},\mathbf{p}\right) \mathcal{M}^{K\rightarrow n}_{H}\left(\mathbf{k}\right)}{2\left(1-e^{a\Delta_{n}^{a\vphantom{b}}}\right)}\left[1+e^{a\Delta_{n}^{a\vphantom{b}}}\left(1-2e^{\Delta_{n}^{a\vphantom{b}}T_{a\vphantom{b}}}\right)\right]+\nonumber \\
&\: a\sum_{m}\dfrac{1}{2E_{m}}\dfrac{\mathcal{M}^{\pi\rightarrow m}_{H}\left(\mathbf{p}\right)\mathcal{M}_{\mu}^{J,K\rightarrow m}\left(\mathbf{k},\mathbf{p}\right)}{2\left(1-e^{-a\Delta_{m}^{b}}\right)}\left[1+e^{-a\Delta_{m}^{b}}\left(1-2e^{-\Delta_{m}^{b}T_{b}}\right)\right], \label{eq:int_spec_rep}
\end{align}
where we define $\mathcal{M}^{J,P_1\rightarrow P_2}_{\mu}\left(\mathbf{k},\mathbf{p}\right)=\left\langle P_2,\mathbf{p}|J_{\mu}|P_1,\mathbf{k}\right\rangle$ and  $\mathcal{M}^{P_1\rightarrow P_2}_{H}\left(\mathbf{p}\right)=\left\langle P_2,\mathbf{p}|H_W|P_1,\mathbf{p}\right\rangle$. To extract the matrix element we must therefore remove the exponentially growing contributions as they appear in the above equation. We remark that one can check explicitly using Eq. (\ref{eq:int_spec_rep}) to show that the matrix element is free of $\mathcal{O}\left(a\right)$ terms. In this exploratory study we perform the simulation with unphysically heavy pions and kaons satisfying $E_K\left(\mathbf{k}\right)<2 M_\pi$, such that the only intermediate state which will give an exponentially growing contribution to the integral consists of a single-pion.

\subsection{Single-pion intermediate state}

Our exploratory simulations use a pion mass of $\sim\unit[430]{MeV}$ and a kaon mass of $\sim\unit[625]{MeV}$; hence only the single-pion exponentially growing contribution must be removed in our analysis. We will now explain the two methods we use to remove these unphysical contributions and present the corresponding numerical discussion in Secs. \ref{sec:m1_removal} and \ref{sec:m2_removal}. A detailed discussion of the treatment of the exponentially growing $\pi\pi$ and $\pi\pi\pi$ intermediate state contributions can be found in Ref.~\cite{Christ:2015aha}.

The first possibility of removing the single-pion exponential is to reconstruct its analytical form from Eq. (\ref{eq:int_spec_rep}). The exponential contribution is therefore
\begin{align}
D_{\mu}^{\pi}\left(T_{a},\mathbf{k},\mathbf{p}\right)=\:a\,\dfrac{1}{2E_{\pi}\left(\mathbf{k}\right)}\dfrac{\mathcal{M}_{\mu}^{J,\pi\to\pi}\left(\mathbf{k},\mathbf{p}\right)\mathcal{M}_{H}^{K\to\pi}\left(\mathbf{k}\right)}{1-e^{-a\Delta^{a}_{\pi}}}e^{\Delta_{\pi}^{a}T_{a}}.\label{eq:single_pion_exp}
\end{align}
The necessary matrix elements and energies can be readily obtained from fits to 2pt and 3pt correlators. We will refer to this method of subtraction as "method 1".

A second method ("method 2") of removing the exponentially growing contribution of the single-pion state is to employ a shift of the weak Hamiltonian by the scalar density, $\bar{s}d$~\cite{Bai:2014cva}.
We choose a constant $c_{s}$ such that
\begin{equation}\label{eq:sd_shift}
\left\langle \pi\left(\mathbf{k}\right)|H_{W}^{\prime}|K\left(\mathbf{k}\right)\right\rangle =\left\langle \pi\left(\mathbf{k}\right)|H_{W}-c_{s}\bar{s}d|K\left(\mathbf{k}\right)\right\rangle=0.
\end{equation}
If we replace $H_{W}$ by $H_{W}^{\prime}$ in Eq. (\ref{eq:int_spec_rep}), the contribution of the single-pion intermediate state vanishes. We can show~\cite{Christ:2015aha} that this shift leaves the total amplitude invariant using the chiral Ward identity
\begin{equation}
i\left(m_{s}-m_{d}\right)\bar{s}d=\partial_{\mu}V_{\bar{s}d}^{\mu}.
\end{equation}

The parameter $c_{s}$ is extracted from the ratio of 3pt correlation functions 
\begin{align}
c_{s}\left(\mathbf{k}\right)=\dfrac{\Gamma^{(3)}_{H_{W}}\left(\mathbf{k}\right)}{\Gamma^{(3)}_{\bar{s}d}\left(\mathbf{k}\right)},
\label{eq:c_s_ratio}
\end{align}
in the region $t_{K}\ll t_{\mathcal{O}} \ll t_{\pi}$, where $t_{\mathcal{O}}$ is the position at which the operator $\mathcal{O}=H_{W}$ or $\bar{s}d$ is inserted. Equivalently $c_s$ may be extracted from the ratio of similar 4pt functions in the region $t_K\ll t_{H} \ll t_{J}$, where we may assume that the 4pt functions are dominated by the exponentially growing contribution of the single-pion intermediate state.

\section{\label{sec:Simulation}Details of the Simulation}

This exploratory study was performed using a $24^{3}\times64$ lattice with an inverse lattice spacing of $1/a=\unit[1.78]{GeV}$, employing Shamir domain wall fermions~\cite{Shamir:1993yf} with Iwasaki gauge action~\cite{Iwasaki:2011np}, a pion mass of $\sim\unit[430]{MeV}$ and a kaon mass of $\sim\unit[625]{MeV}$~\cite{Aoki:2010dy,Blum:2014tka}. We use a sample of 128 configurations, each separated by 20 molecular dynamics time units. In order to cancel divergences with the GIM mechanism we include a charm quark with a bare mass of $am=0.2$. Using the mass renormalization factor $Z_{m}^{\MSbar}(2\,\mathrm{GeV})=1.498$ for this lattice~\cite{Aoki:2010dy}, this corresponds to an unphysical charm quark of mass $m_c^{\overline{\textrm{MS}}}(2\,\textrm{GeV})=533\,\mathrm{MeV}$.

The renormalization of the $H_{W}$ operator is simplified considerably by our prescription of domain wall fermions: the good chiral symmetry prevents the mixing of the operators $Q_1$ and $Q_2$ (from Eq.~(\ref{eq:weak_operator})) with right-handed operators. The details of the nonperturbative renormalization of this operator are given in Ref.~\cite{Christ:2012se}, where the ensembles used to perform the nonperturbative renormalization have the same lattice spacing and action, but a smaller volume. The results are also valid for our lattice as the renormalization procedure depends upon the UV behavior of the theory and thus is insensitive to finite volume effects.

We now move to a detailed explanation of the setup of our calculation. In the next subsection we will introduce the schematic of the relevant 4pt correlator and give an overview of the propagators we choose to use to perform each of the contractions involved in the construction of the correlator. In Sec.~\ref{sec:imp_details} we will give a more technical discussion of the implementation.

\subsection{\label{sec:calc_setup}Setup of the calculation}

We simulate a kaon with momentum $\mathbf{k}=0$ at a time $t_K=0$ decaying into a pion with momentum $\mathbf{p}$ at $t_{\pi}=28$. We have considered three separate final state pion momenta: $\mathbf{p}=\frac{2\pi}{L}(1,0,0)$, $\mathbf{p}=\frac{2\pi}{L}(1,1,0)$ and $\mathbf{p}=\frac{2\pi}{L}(1,1,1)$, where $L=24$ is the spatial extent of our lattice. We will thus label each kinematical case by the momentum $\mathbf{p}$. In all cases the current is situated halfway between the kaon and pion at $t_J=14$; this position is chosen such that we can integrate over $t_H$ in a window around the current and be far enough away from the positions of the pion/kaon interpolators to avoid the contamination of excited state contributions. We use Coulomb gauge-fixed wall sources in our calculation to give good overlap with the ground state pion and kaon, which allows us to keep the kaon-pion source-sink separation as small as possible to achieve the best possible signal for the amplitude.

\begin{figure}
	\begin{center}
		\begin{tabular}{c c}
			\includegraphics[width=0.4\linewidth]{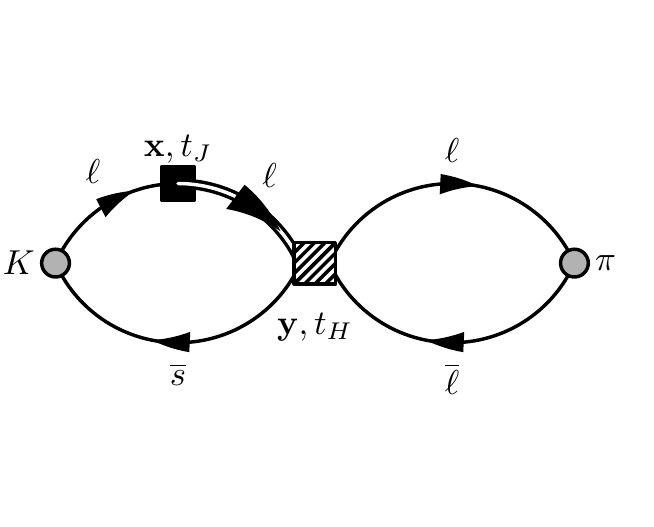} &
			\includegraphics[width=0.4\linewidth]{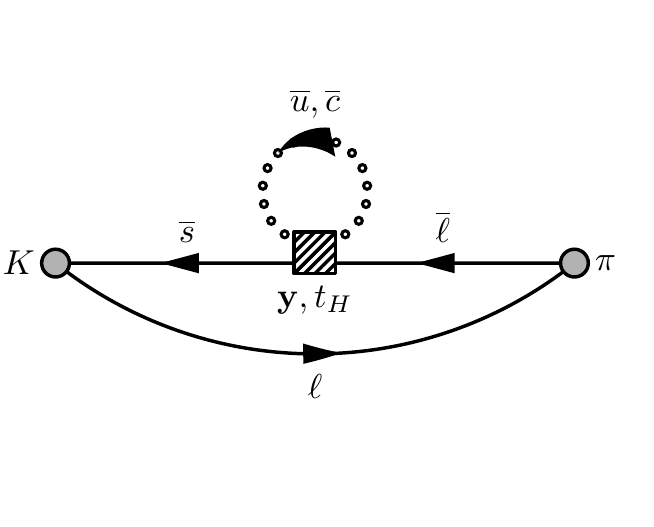} \\
		\end{tabular}
	\end{center}
	\caption{\label{fig:prop_diag}Demonstration of how propagators are used to construct diagrams. The position of the $H_W$ operator is indicated by the shaded square, and may be placed at any spacetime position. The insertion of the current is denoted by a black square, fixed on an single time slice and summed over space. The double line represents the part of the propagator computed using a sequential inversion; the dotted line represents the loop propagator, computed using spin-color diluted random volume sources~\cite{Bernardson:1993he}.}
\end{figure}

The computation of the full set of diagrams corresponding to the rare kaon decay can be accomplished by computing 14 propagators. Four are required to connect the kaon/pion sources to the $H_W$ insertion: one strange and one light for the kaon; two light propagators with momenta $\mathbf{0}$ and $\mathbf{p}$ to produce a pion with momentum $\mathbf{p}\ne\mathbf{0}$ (this also allows us to make a pion with momentum $\mathbf{0}$). Two more propagators are needed for the loops in the $S$ and $E$ and disconnected diagrams (one light, one charm), and one more for the strange loop in disconnected diagrams. We use each of these seven propagators to calculate a sequential propagator to achieve the current insertion to bring us up to 14. The types of propagators used are shown schematically in Fig.~\ref{fig:prop_diag}. Furthermore, to construct all the 2pt and 3pt functions required for our analysis procedure, we also compute one additional strange propagator with momentum $\mathbf{p}$ such that we can produce a kaon with momentum $\mathbf{p}$.

\subsection{\label{sec:imp_details}Details of the Implementation}

To compute the loops in the $S$ and $E$ diagrams we require the propagator from each site to that exact same site for each color and spin index, i.e. the diagonal entries of the inverse of the Dirac operator. This is readily accomplished by making use of random spin-color diluted volume sources~\cite{Bernardson:1993he,Foster:1998vw,McNeile:2006bz}; the details of these sources are discussed in Appendix~\ref{sec:random_vol}. With such a propagator the position of the $H_W$ operator can be inserted at any position on the lattice, thus enabling the integration of the position of $H_W$ over the whole lattice.

The insertions of the electromagnetic current can be achieved using sequential propagators, with the current inserted at a time $t_J$. We only consider the element $\mu=0$ of the current to save computational resources, which is enough to extract the form factor using Eq. (\ref{eq:mat_elem_form_fac}). The computation of sequential propagators is discussed in Appendix~\ref{sec:seq_prop}. With the current fixed at a single time the time ordering of the operators is straightforward to implement, which simplifies our analysis procedure. Another advantage is that the current is automatically summed over the entire spatial volume. For our lattice this spatial sum reduces the statistical error by approximately a factor of 3. The primary disadvantage of this method is that we must perform a new set of inversions if we wish to consider the current at another temporal position, with a different initial (final) state momentum of the kaon (pion) or for a different polarization.

In our present calculation we omit the disconnected diagrams where the electromagnetic current is self-contracted (see Fig.~\ref{fig:J_contractions}). The primary reason for this is practical: we expect the disconnected contribution to be very noisy and thus would require a significantly larger statistical sample to be measured to obtain a signal comparable to the other diagrams (relative to noise). However we also  expect the disconnected contribution to be suppressed by a factor of $1/{N_c}$ and by the  approximate $SU(3)$ flavor symmetry. In the continuum we would expect the disconnected contribution to have $\sim 10\%$ of the contribution of the connected part~\cite{DellaMorte:2010aq}. With our choice of masses the $SU(3)$ suppression is stronger and so the disconnected diagrams are expected to be further suppressed. Nevertheless, our simulation is set up such that the disconnected contribution can be calculated separately to the connected contributions, and can be added at a later stage without having to recalculate any propagators or the connected diagrams that we have already.

For our simulation we choose to use $N_{\eta}=14$ random noise sources on each configuration to obtain a reasonable signal for the loop function of the $S$ and $E$ diagrams. While increasing $N_{\eta}$ would increase our precision further, we found $N_{\eta}=14$ to be a reasonable compromise when also taking into account available computational resources. In addition to this we translate the computation of the 4pt correlator to $N_t=12$ positions over the time direction of our lattice on a single configuration. Each translation ultilizes the same noise propagators generated for the loop diagrams; however we find the signal-to-noise ratio of the $S$ and $E$ diagrams increases by approximately a factor of 3 when we include these additional translations. This is consistent with the increase in statistical precision expected if the translations are statistically independent of each other.

We chose time positions for the operators in this decay such that there exists a large enough window to fully integrate over $t_H$ on either side of the current. In such a setup, we found that the closer the position of the current to the pion, the better the signal for the decay. We therefore tested simulating with an additional time position for the current placed closer to the pion such that we may integrate over the region $\left[t_{J}-T_{a},t_{J}\right]$ with an improved precision. We found that this second current insertion would increase the simulation cost by $\sim 50\%$, but reduce the statistical error by a factor of $\sim 25\%$. However the additional cost of these inversions means that the decay can only be translated across eight time positions in the same amount of CPU time as it costs to perform 12 translations with a single current position. We found that the loss of precision from considering fewer translations ultimately canceled the increase from the second current position.

\begin{table}
	\begin{centering}
		\begin{tabular}{|c|c|c|c|c|}
			\hline 
			\multirow{2}{*}{Description} & \multirow{2}{*}{Source Type} & \multicolumn{3}{c|}{Number of Inversions}\tabularnewline
			\cline{3-5} 
			&  & Light & Strange & Charm\tabularnewline
			\hline 
			\hline 
			$C$ and $W$ propagators & Gauge-fixed wall & $3N_{t}$ & $N_{t}$ & 0\tabularnewline
			\hline 
			$S$ and $E$ loops & Random volume & $N_{\eta}$ & 0 & $N_{\eta}$\tabularnewline
			\hline 
			Current insertions & Sequential & $\left(3+N_{\eta}\right)N_{t}$ & $N_{t}$ & $N_{\eta}N_{t}$\tabularnewline
			\hline 
			Analysis supplements & Gauge-fixed wall & $0$ & $N_{t}$ & 0\tabularnewline
			\hline 
			Total & - & $N_{\eta}+N_{t}\left(6+N_{\eta}\right)$ & 3$N_{t}$ & $N_{\eta}\left(N_{t}+1\right)$\tabularnewline
			\hline 
			$N_{\eta}=14$, $N_{t}=12$ & - & 254 & 36 & 182\tabularnewline
			\hline 
		\end{tabular}
		\par\end{centering}
	
	\caption{\label{tab:prop_summary}Summary of propagators calculated in our simulation for a single choice of pion momentum on a single configuration, and the corresponding number of inversions required. $N_{\eta}$ is the number of noise vectors used in the computation of the quark loops; $N_{t}$ is the number of translations in the time direction across a single configuration at which all the contractions are computed.}
\end{table}

On a single configuration we thus require 254 light propagator inversions, 36 strange inversions and 182 charm inversions (including disconnected diagrams would require a further 182 strange inversions). This is summarized in Table~\ref{tab:prop_summary}. Because of this large number of light propagators we made use of the HDCG algorithm~\cite{Boyle:2014rwa} to accelerate the light-quark inversions. The overhead of deflating the Dirac operator costs the equivalent of two to three conjugate gradient (CG) inversions; however the cost of a light-quark inversion is subsequently reduced by a factor of 4.

\section{\label{sec:Results}Numerical Results}

Ultimately the aim of this calculation is to demonstrate that the matrix element of $K\rightarrow\pi\gamma^{*}$ decays can be determined with controlled systematic errors. In this section we will discuss the numerical results of our simulation, and include a critical discussion of our two primary analysis techniques. For demonstration purposes we will focus on the results for our kinematics with a charged kaon at rest decaying into a charged pion with one unit of momentum in one spatial direction. 

While it is also possible to compute the neutral decay $K_S\rightarrow \pi^0 \ell^+ \ell^-$ using our lattice data, with our current statistics we find that we do not obtain any worthwhile signal for this correlator. The error is dominated by the additional, disconnected contractions shown in Fig.~\ref{fig:pi0_contractions}. These contractions appear much noisier than the other diagrams, and their error is many times larger than the signal from the remaining contractions. The difficulty to extract a signal from our data can also be understood physically: we have only considered photons with small momenta; the structure of the kaon/meson is thus not well enough resolved to obtain a clear signal. When we simulate the decay into a pion with a higher momentum the structure is better resolved, although the correlators with momentum are naturally more noisy. This makes them difficult to analyze with the size of our present statistical sample. For this reason we will focus exclusively on the charged channel, and will discuss the neutral channel in later works.

\subsection{Lattice correlators}

\begin{figure}
	\begin{center}
		\begin{tabular}{c c}
			\includegraphics[width=0.5\textwidth]{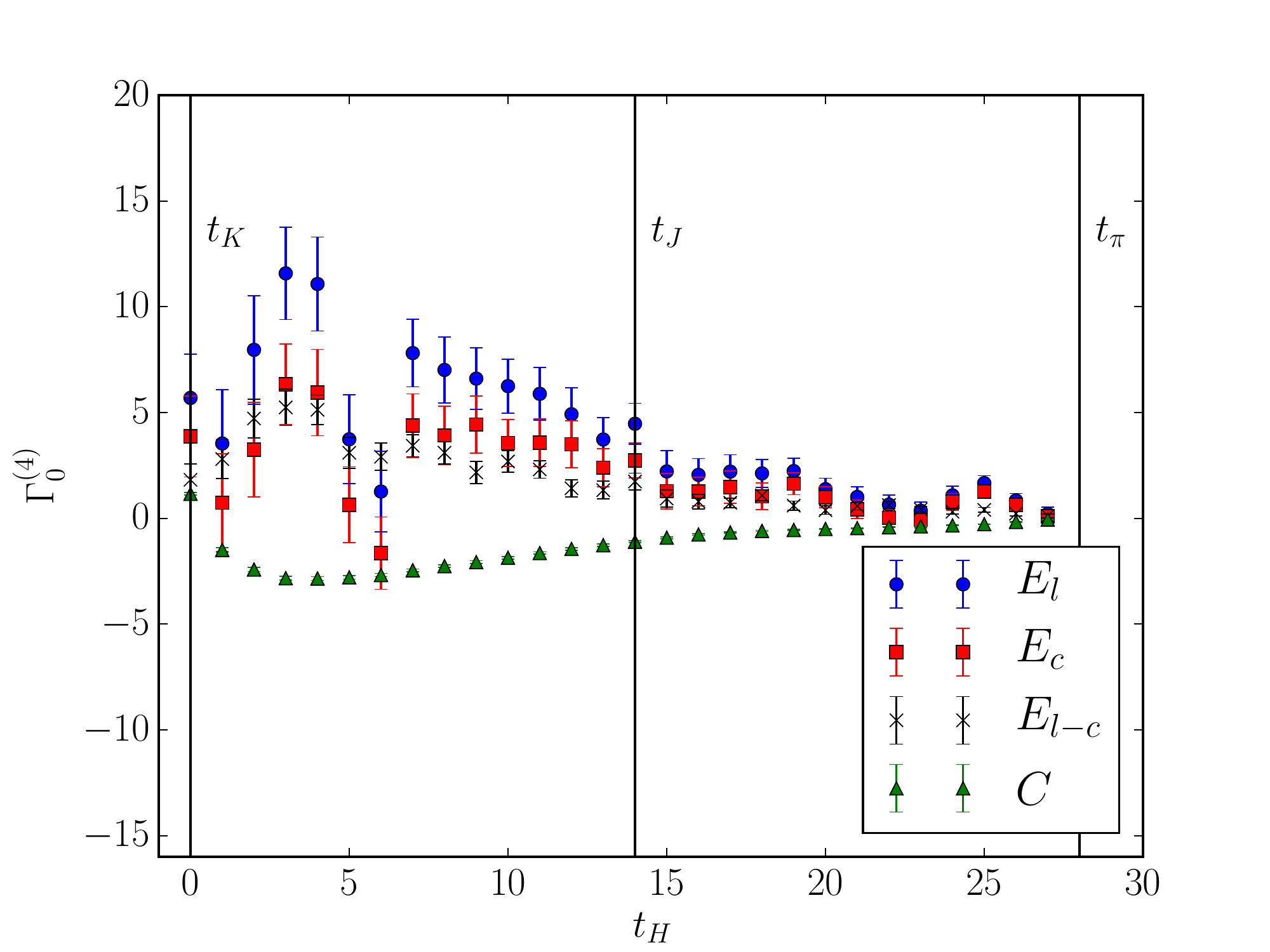} &
			\includegraphics[width=0.5\textwidth]{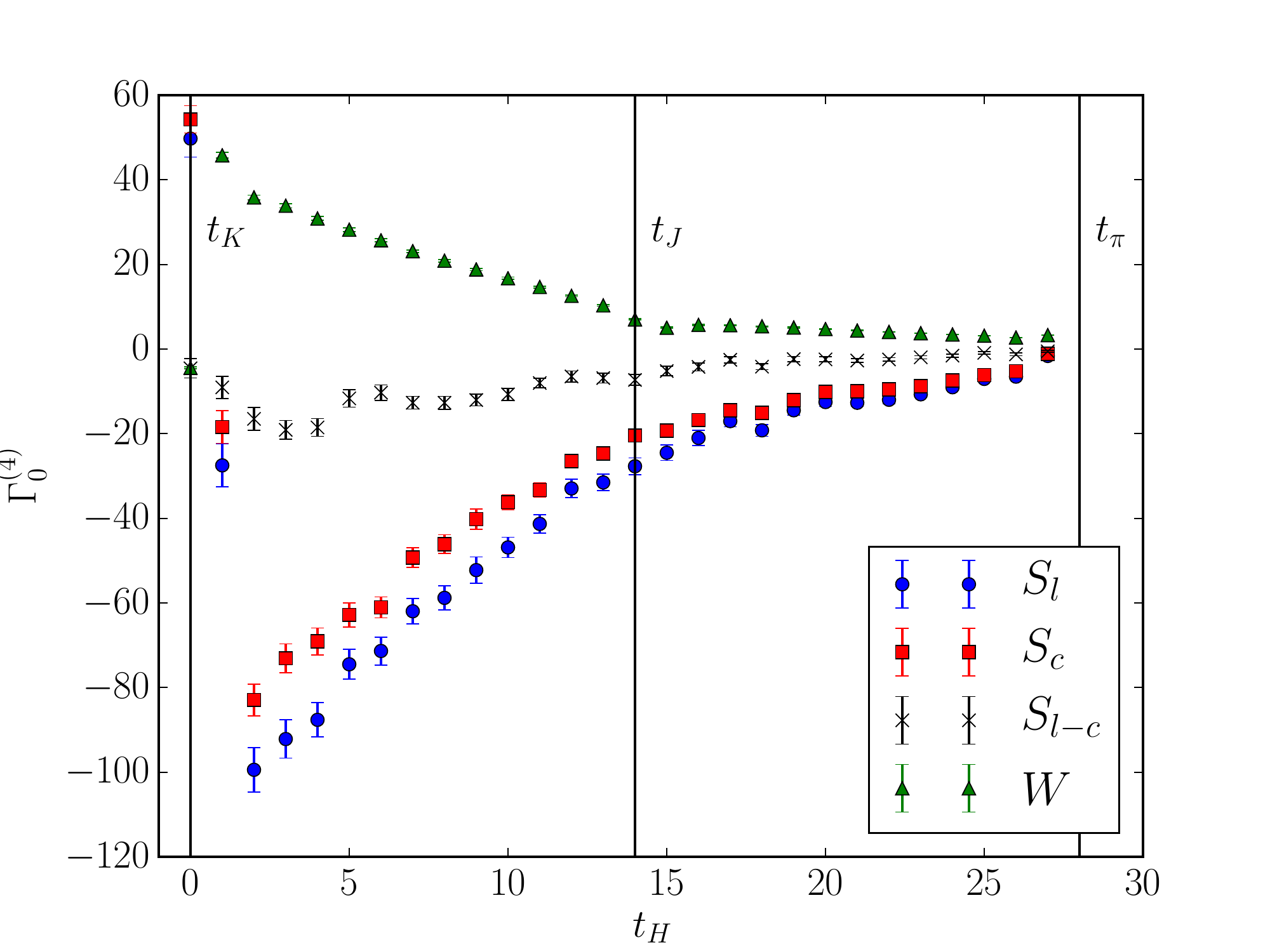} \\
			(a) & (b) \\
		\end{tabular}
	\end{center}
	\caption{\label{fig:GIM_subtraction}The contributions of each of the diagrams to the rare kaon decay corresponding to the weak operators (a) $Q_1$ and (b) $Q_2$, both before and after the GIM subtraction. Each diagram has been constructed using the appropriate fractional quark charges (excluding the overall charge factor $e$), and the correlators have been multiplied by the relevant renormalization constants and Wilson coefficients for matching to the $\MSbar$ scheme (as described in detail in Ref.~\cite{Christ:2012se}). Time positions of the kaon/pion interpolators and current insertion are indicated.}
\end{figure}

In Fig.~\ref{fig:GIM_subtraction} we show the contributions of each of the diagrams to the 4pt correlator that correspond to the charged rare kaon decay. A comparison of Fig.~\ref{fig:GIM_subtraction} $(a)$ and $(b)$ shows that the dominant contribution to the decay comes from the $Q_2$ operator, i.e. the $W$ and $S$ diagrams. Furthermore as the loop diagrams $S$ and $E$ are considerably noisier than $W$ and $C$, it follows that the $S$ diagram will dominate the error on our final result. We remark that each diagram in Fig.~\ref{fig:GIM_subtraction} has already been multiplied by the appropriate renormalization constants to match to the $\MSbar$ scheme, as defined in Table V of Ref.~\cite{Christ:2012se}. For the scale $\mu=2.15\,\mathrm{GeV}$, we thus multiply our bare lattice operators $Q_1$ and $Q_2$ by the coefficients $C_{1}^{\lat}=-0.2216$ and $C_{2}^{\lat}=0.6439$ respectively. For this analysis we neglect any systematic errors on these Wilson coefficients, as they are not a primary concern of our exploratory studies. However, a full discussion of systematic errors of the renormalization of the $H_W$ operator has previously been given in the context of $K\to\pi\pi$ decays; see e.g. Refs.~\cite{Blum:2011pu,Blum:2015ywa}.

\begin{figure}
	\begin{center}
		\begin{tabular}{c c}
			\includegraphics[width=0.5\textwidth]{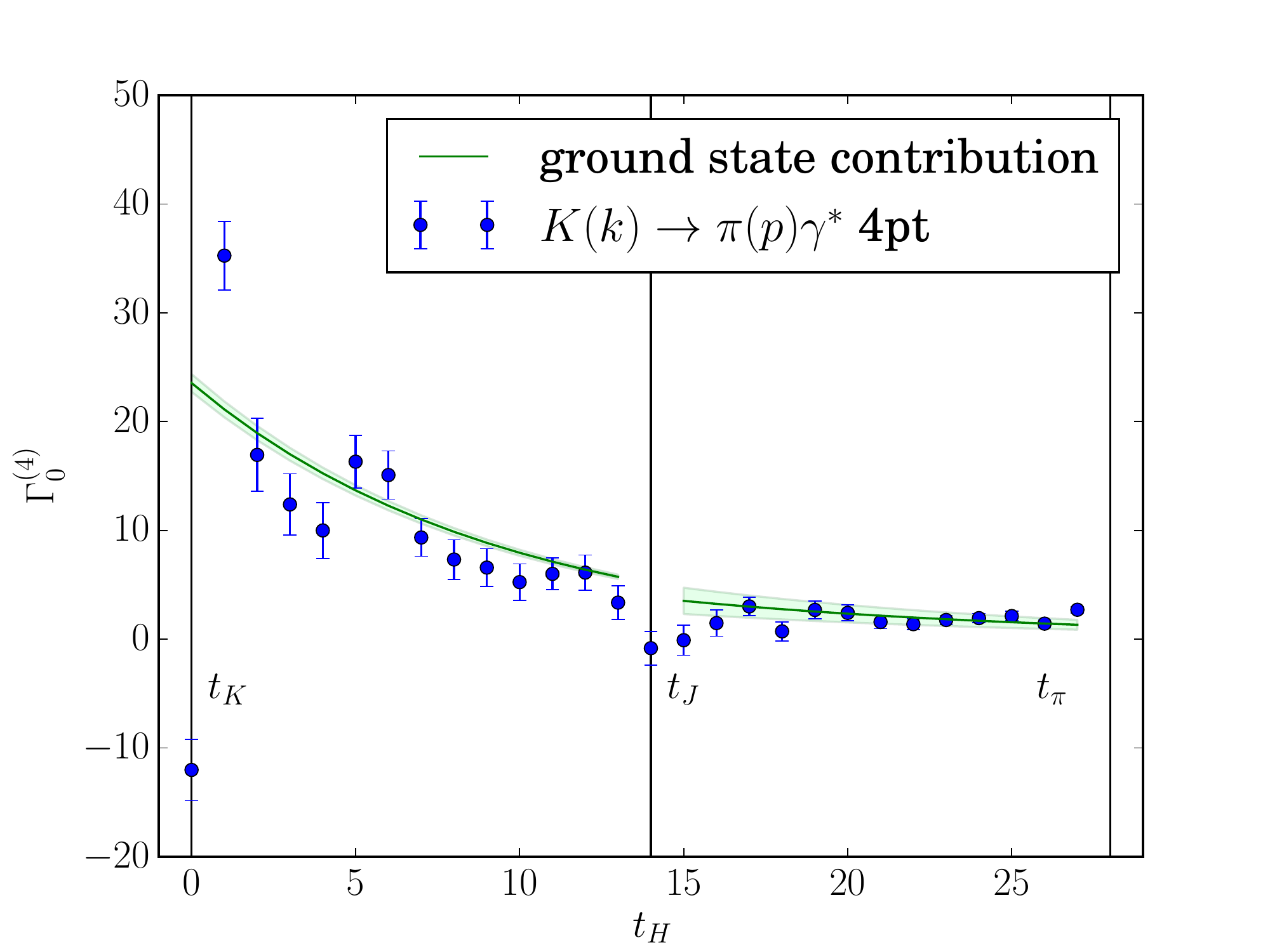} &
			\includegraphics[width=0.5\textwidth]{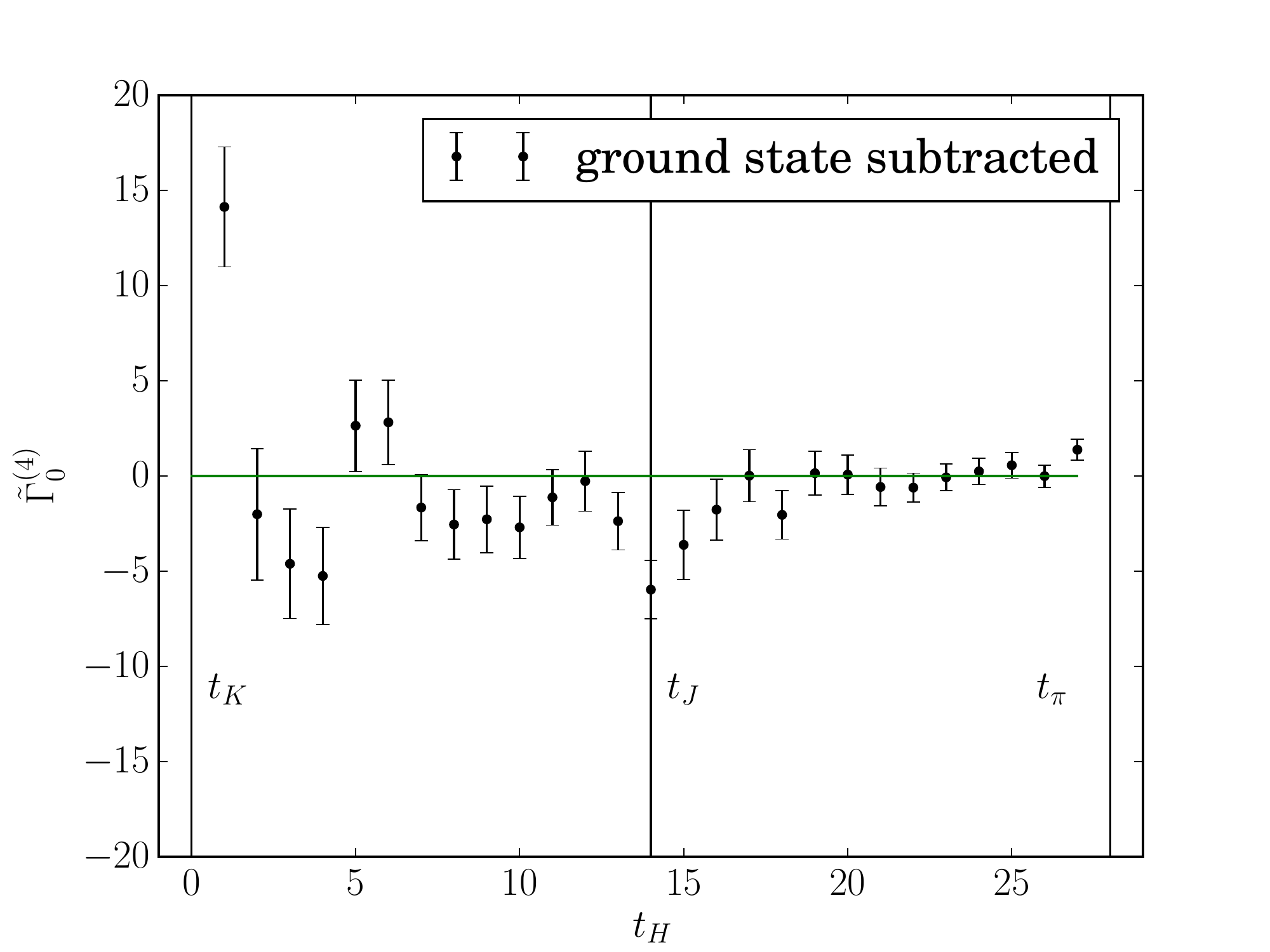} \\
			(a) & (b)
		\end{tabular}
	\end{center}
	\caption{\label{fig:4pt_correlator}(a) The 4pt rare kaon decay correlator measured in our simulation with $\textbf{k}=(0,0,0)$ and $\textbf{p}=\frac{2\pi}{L}(1,0,0)$. The ground state contribution has been constructed from fits to 2pt and 3pt correlators. (b) The 4pt correlator after removing the ground state contribution (i.e. the single-pion and single kaon intermediate states). Time positions of the kaon/pion interpolators and the current insertion are indicated.}
\end{figure}

Additionally in Fig.~\ref{fig:GIM_subtraction} we show how the $S$ and $E$ diagrams are obtained by subtracting the charm loop diagram from the up quark loop diagram, i.e. the GIM subtraction. Here we expect the GIM subtraction to be more severe than in the physical case, as we are using a lighter-than-physical charm quark and a heavier-than-physical light quark. With physical masses we should expect the $S$ diagram to have a larger magnitude. In the final correlator the $S$ and $W$ diagrams appear to add destructively; this may have a severe effect on the final result if there is a large degree of cancellation between the contributions of the $S$ and $W$ diagrams to the final matrix element. The combined rare kaon decay 4pt correlators that we analyse are shown in Fig.~\ref{fig:4pt_correlator}. We show these correlators before and after the removal of unphysical exponential terms that appear as a relic of the Euclidean formulation~\cite{Christ:2015aha}. The removal of these terms is discussed in the following section.

\subsection{\label{sec:m1_removal}Removal of single-pion exponential: Method 1}

The main difficulty of this analysis is the removal of the exponential term that grows with increasing $T_{a}$; however in practice we find that it is necessary to also consider the term that falls exponentially with $T_{b}$ [as seen in the second line of Eq.~(\ref{eq:int_spec_rep})], as the integral does not converge in the available time extent. This can be attributed to the fact that here the kaon-pion mass difference is rather small; hence the exponent for the exponential decay is small. In practice therefore it is necessary also to remove the single kaon contribution that decays with $T_{b}$ in a manner similar to the exponentially growing term by reconstructing the state from 2pt/3pt functions. Asymptotically the integrated 4pt correlator can be written in the form:
\begin{align}
I_{\mu}\left(T_{a},T_{b},\mathbf{k},\mathbf{p}\right)=&A_{\mu}\left(\mathbf{k},\mathbf{p}\right)+c_{\mu}^{1}\left(\mathbf{k},\mathbf{p}\right)e^{\Delta_{\pi}^{a}T_{a}}\left[\dfrac{\Delta_{\pi}^{a}}{1-e^{-\Delta_{\pi}^{a}}}\right]\nonumber\\
& \phantom{\mathcal{A}_{\mu}\left(\mathbf{k},\mathbf{p}\right)}+ c_{\mu}^{2}\left(\mathbf{k},\mathbf{p}\right)e^{-\Delta_{K}^{b}T_{b}}\left[\dfrac{\Delta_{K}^{b}}{e^{\Delta_{K}^{b}}-1}\right]+\ldots,
\label{eq:asymptotic_form}
\end{align}
with $\Delta_{\pi}^{a}=E_{K}\left(\mathbf{k}\right)-E_{\pi}\left(\mathbf{k}\right)$ and $\Delta_{K}^{b}=E_{K}\left(\mathbf{p}\right)-E_{\pi}\left(\mathbf{p}\right)$. The terms in the square brackets, which tend towards 1 in the continuum limit, account for the corrections necessary to treat the single meson intermediate states (i.e. the ground state contributions) using a discrete formulation. In terms of particle energies and matrix elements from 3pt functions we can write
\begin{align}
c_{\mu}^{1}\left(\mathbf{k},\mathbf{p}\right)=\dfrac{\mathcal{M}_{\mu}^{J,\pi}\left(\mathbf{k},\mathbf{p}\right)\mathcal{M}_{H}\left(\mathbf{k}\right)}{2E_{\pi}\left(\mathbf{k}\right)\Delta_{\pi\vphantom{K}}^{a\vphantom{b}}},&\quad
c_{\mu}^{2}\left(\mathbf{k},\mathbf{p}\right)=-\dfrac{\mathcal{M}_{\mu}^{J,K}\left(\mathbf{k},\mathbf{p}\right)\mathcal{M}_{H}\left(\mathbf{p}\right)}{2E_{K}\left(\mathbf{p}\right)\Delta_{K}^{b}},
\label{eq:c_1_c_2}
\end{align}
where $\mathcal{M}^{J,P}_{\mu}\left(\mathbf{k},\mathbf{p}\right)=\left\langle P,\mathbf{p}|J_{\mu}|P,\mathbf{k}\right\rangle$ and $\mathcal{M}_{H}\left(\mathbf{k}\right)=\left\langle K\left(\mathbf{k}\right)|H_W|\pi\left(\mathbf{k}\right)\right\rangle$. Our analysis thus proceeds by removing the terms proportional to $c_{\mu}^{1}$ and $c_{\mu}^{2}$ from the 4pt correlator, and fitting the remainder to a constant to obtain $A_{\mu}$, which is the amplitude in Euclidean space, up to a factor as seen in Eq.~(\ref{eq:Euclid_to_Minkowski}).

It is indeed possible to use Eq. (\ref{eq:asymptotic_form}) to fit the 4pt function directly to remove the ground state contributions. Because the exponents can be obtained much more accurately from 2pt functions, we simply fit the parameters $A_{\mu}$, $c_{\mu}^{1}$ and $c_{\mu}^{2}$ in the region where the ground state contributions dominate. We find that we obtain consistent results when we use this procedure.

\begin{figure}
	\begin{center}
		\includegraphics[width=0.65\textwidth]{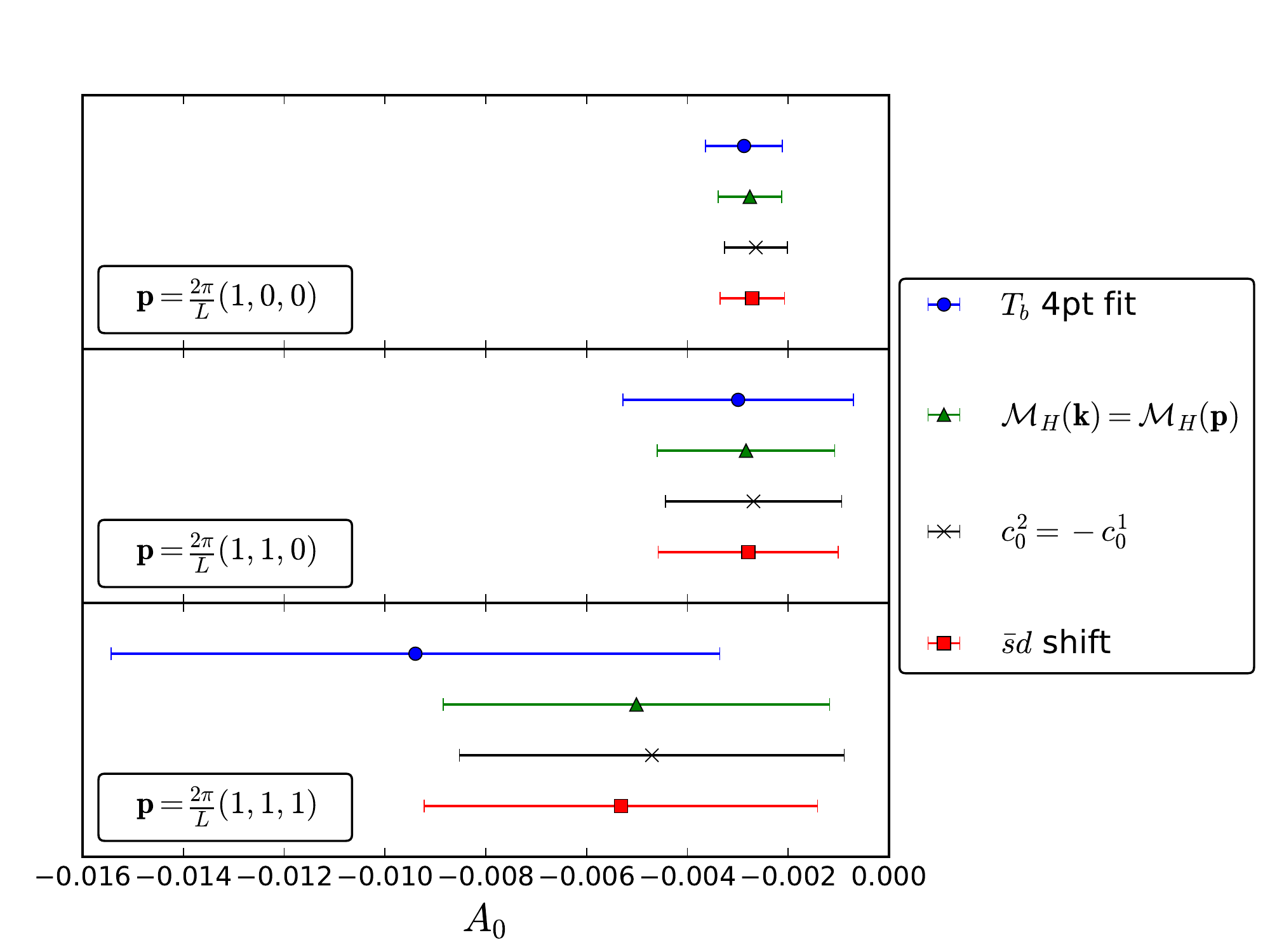}
	\end{center}
	\caption{\label{fig:methods_comparison}Plot of the amplitudes (in lattice units) obtained using each of the different analysis methods.}
\end{figure}

The computed values for the coefficients $c_{0}^{1}$ and $c_{0}^2$ [obtained using both Eq. (\ref{eq:c_1_c_2}) and the direct 4pt fit] are shown in Table~\ref{tab:coeffs}. We remark that the coefficient $c_{0}^2$ becomes significantly less well determined when we increase the momentum of the pion. The reason for this is that the matrix element $\mathcal{M}_{H}\left(\mathbf{p}\right)$ is difficult to determine precisely when we have $\mathbf{p}\ne \mathbf{0}$. We can thus avoid introducing an unnecessarily large statistical error either by fitting $c_{0}^2$ directly from the 4pt correlator or by making well-motivated approximations. The two approximations we have considered are $c_{0}^{2}=-c_{0}^{1}$, and $\mathcal{M}_H\left(\mathbf{k}\right)=\mathcal{M}_H\left(\mathbf{p}\right)$. The first approximation holds exactly when $\mathbf{k}=\mathbf{p}$; the second holds exactly in the $SU(3)$ flavor symmetric limit, i.e. when $M_{\pi}=M_K$. A short proof of each of these statements can be found in Appendix~\ref{sec:c_1_c_2_approx}. A summary of the matrix elements obtained using each of these methods can be found in Table~\ref{tab:results}, and are displayed graphically in Fig.~\ref{fig:methods_comparison}. We remark that the approximations of $c_{0}^{2}$ need not be exact: they are sufficient if the systematic error on the approximation is significantly smaller than the statistical error on the final signal for the amplitude. Taking correlated differences between the different analysis techniques reveals that the systematic errors on these approximations are substantially less than the statistical errors on the matrix elements.

\begin{figure}
	\begin{center}
		\begin{tabular}{c c}
			\includegraphics[width=0.5\textwidth]{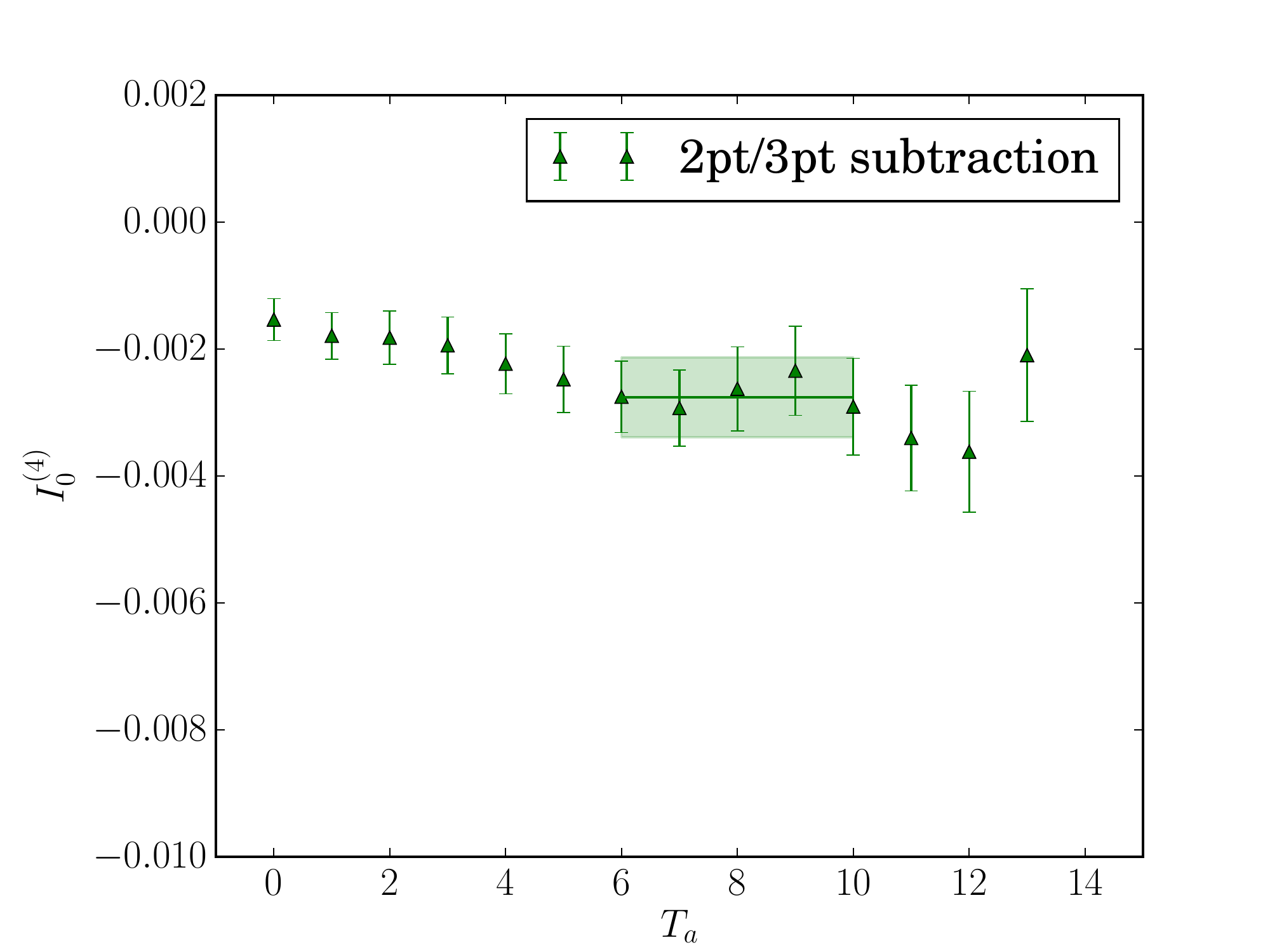} &
			\includegraphics[width=0.5\textwidth]{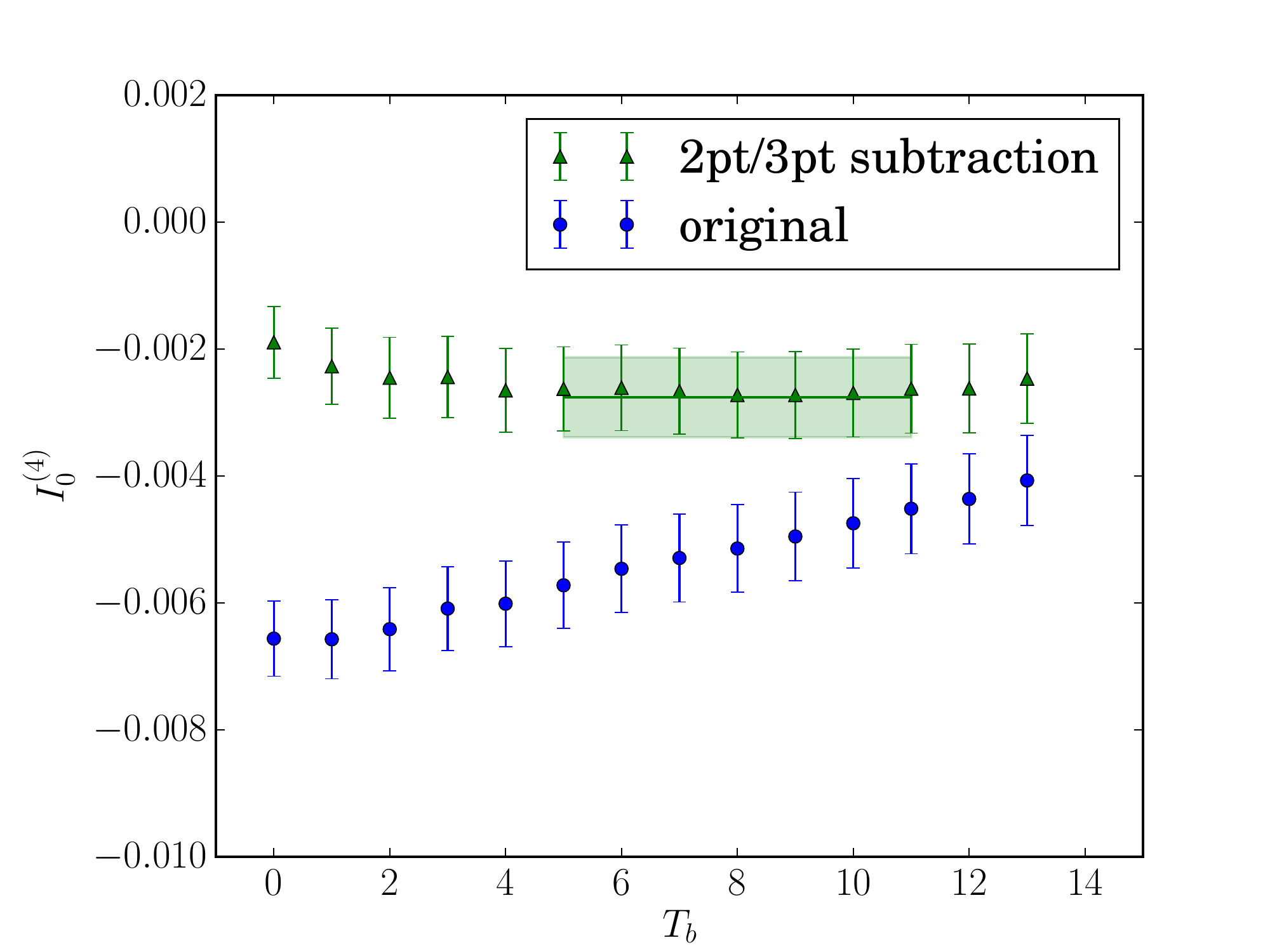} \\
			(a) & (b) \\
		\end{tabular}
	\end{center}
	\caption{\label{fig:method_1}The integrated 4pt correlator, shown for (a) $\int_{t_J-T_a}^{t_J+8}\tilde{\Gamma}^{(4)}_{0}dt_H$ to demonstrate the $T_a$ dependence and (b) $\int_{t_J-6}^{t_J+T_b}\tilde{\Gamma}^{(4)}_{0}dt_H$ to demonstrate the $T_b$ dependence. The single-pion exponential growth has been removed using method 1. The single kaon exponential decay has been removed using the approximation $\mathcal{M}_H\left(\mathbf{p}\right)=\mathcal{M}_H\left(\mathbf{k}\right)$. The position of the plateau corresponds to $A_{0}=-0.0028(6)$ obtained by a fit to the data over the indicated range.}
\end{figure}

In Fig.~\ref{fig:method_1} (a) we display the $T_a$ and in Fig.~\ref{fig:method_1} (b) the $T_b$ dependence of the integrated 4pt correlator having removed the ground state contributions. In Fig.~\ref{fig:method_1}(a) we see that after the analytic removal of the single-pion intermediate state, no other exponentially growing states are discernible beyond statistical errors. This suggests that contributions from excited states are adequately suppressed. Fig.~\ref{fig:method_1}(b) demonstrates the slow exponential decay in $T_b$ which is caused by the small exponent $E_{K}(\mathbf{p})-E_{\pi}(\mathbf{p})$. This appears to be a problem only because our pion and kaon masses are unphysically close; in simulations closer to the physical masses the exponent $E_{K}(\mathbf{p})-E_{\pi}(\mathbf{p})$ will become larger; hence the residual $T_b$ dependence will decay more quickly. Consequently this subtraction may become unnecessary in future studies, although in any case it can be removed as we have shown here.

\begin{table}
	\begin{centering}
		\begin{tabular}{|c|c|c|c|c|}
    \hline
     \multirow{2}{*}{Coefficient} & \multirow{2}{*}{Description} & \multicolumn{3}{c|}{Kinematic} \tabularnewline
    \cline{3-5}
     & & $\mathbf{p}=\frac{2\pi}{L}(1,0,0)$ & $\mathbf{p}=\frac{2\pi}{L}(1,1,0)$ & $\mathbf{p}=\frac{2\pi}{L}(1,1,1)$ \tabularnewline
    \hline
     \multirow{2}{*}{$c^{1}_{0}(\mathbf{k}, \mathbf{p})$} & 4pt fit & 0.00523(45) & 0.0056(13) & 0.0050(33) \tabularnewline
     & 2pt/3pt & 0.00538(18) & 0.00549(20) & 0.00611(32) \tabularnewline
    \hline
     \multirow{3}{*}{$c^{2}_{0}(\mathbf{k}, \mathbf{p})$} & $\mathcal{M}_H(\mathbf{p})=\mathcal{M}_H(\mathbf{k})$ & -0.00487(18) & -0.00494(22) & -0.00532(48) \tabularnewline
     & 4pt fit & -0.00464(62) & -0.0046(22) & 0.0012(56) \tabularnewline
     & 2pt/3pt & -0.0050(17) & -0.025(20) & 0.06(12) \tabularnewline
    \hline
     \multirow{3}{*}{$c^{1}_{0}(\mathbf{k}, \mathbf{p})+c^{2}_{0}(\mathbf{k}, \mathbf{p})$} & $\mathcal{M}_H(\mathbf{p})=\mathcal{M}_H(\mathbf{k})$ & 0.000516(44) & 0.00055(12) & 0.00079(38) \tabularnewline
     & 4pt fit & 0.00075(61) & 0.0009(22) & 0.0073(56) \tabularnewline
     & 2pt/3pt & 0.0004(17) & -0.020(20) & 0.06(12) \tabularnewline
    \hline
\end{tabular}

		\par\end{centering}
	
	\protect\caption{\label{tab:coeffs}Parameters of Eq. (\ref{eq:asymptotic_form}) (in lattice units) obtained via analytic reconstruction using 2pt and 3pt fit results or fitting the integrated 4pt correlator directly. For $c_{0}^{2}$ the result using the approximation $\mathcal{M}_{H}\left(\mathbf{p}\right)=\mathcal{M}_{H}\left(\mathbf{k}\right)$ is also shown.}
	
\end{table}

\begin{table}
	\begin{centering}
		\begin{tabular}{|c|c|c|c|c|}
    \hline
     Analysis & Kinematic & $A_{0}$ & $A_{0}^{C,W}$ & $A_{0}^{S,E}$ \tabularnewline
    \hline
     \multirow{3}{*}{\makecell{method 1\\($\mathcal{M}_H(\mathbf{p})=\mathcal{M}_H(\mathbf{k})$)}} & $\mathbf{p}=\frac{2\pi}{L}(1,0,0)$
& -0.00276(63) & -0.00161(14) & -0.00106(60) \tabularnewline
     & $\mathbf{p}=\frac{2\pi}{L}(1,1,0)$
& -0.0028(18) & -0.00251(40) & -0.0003(17) \tabularnewline
     & $\mathbf{p}=\frac{2\pi}{L}(1,1,1)$
& -0.0050(38) & -0.0027(12) & -0.0023(39) \tabularnewline
    \hline
     \multirow{3}{*}{\makecell{method 1\\($c_{0}^{2}=-c_{0}^{1}$)}} & $\mathbf{p}=\frac{2\pi}{L}(1,0,0)$
& -0.00264(62) & -0.00133(12) & -0.00122(60) \tabularnewline
     & $\mathbf{p}=\frac{2\pi}{L}(1,1,0)$
& -0.0027(17) & -0.00217(33) & -0.0005(17) \tabularnewline
     & $\mathbf{p}=\frac{2\pi}{L}(1,1,1)$
& -0.0047(38) & -0.00196(84) & -0.0028(39) \tabularnewline
    \hline
     \multirow{3}{*}{\makecell{method 1\\(4pt fit)}} & $\mathbf{p}=\frac{2\pi}{L}(1,0,0)$
& -0.00288(76) & -0.00169(17) & -0.00109(73) \tabularnewline
     & $\mathbf{p}=\frac{2\pi}{L}(1,1,0)$
& -0.0030(23) & -0.00298(52) & -0.0000(22) \tabularnewline
     & $\mathbf{p}=\frac{2\pi}{L}(1,1,1)$
& -0.0094(60) & -0.0041(13) & -0.0053(61) \tabularnewline
    \hline
     \multirow{3}{*}{method 2} & $\mathbf{p}=\frac{2\pi}{L}(1,0,0)$
& -0.00271(64) & -0.00151(16) & -0.00110(58) \tabularnewline
     & $\mathbf{p}=\frac{2\pi}{L}(1,1,0)$
& -0.0028(18) & -0.00240(48) & -0.0004(17) \tabularnewline
     & $\mathbf{p}=\frac{2\pi}{L}(1,1,1)$
& -0.0053(39) & -0.0034(12) & -0.0019(38) \tabularnewline
    \hline
     \multirow{3}{*}{$c_s\times\bar{s}d$} & $\mathbf{p}=\frac{2\pi}{L}(1,0,0)$
& -0.000010(84) & -0.00002(20) & 0.00001(11) \tabularnewline
     & $\mathbf{p}=\frac{2\pi}{L}(1,1,0)$
& -0.00002(21) & -0.00005(49) & 0.00003(28) \tabularnewline
     & $\mathbf{p}=\frac{2\pi}{L}(1,1,1)$
& 0.00032(52) & 0.0007(12) & -0.00042(69) \tabularnewline
    \hline
\end{tabular}

		\par\end{centering}
	\protect\caption{\label{tab:results}Summary of matrix elements obtained using various analysis methods. All values are given in lattice units. Results are shown for all classes of diagrams, and also separated into the nonloop and loop contributions.}
\end{table}

\subsection{\label{sec:m2_removal}Removal of single-pion exponential: Method 2}

\begin{figure}
	\begin{center}
		\includegraphics[width=0.5\textwidth]{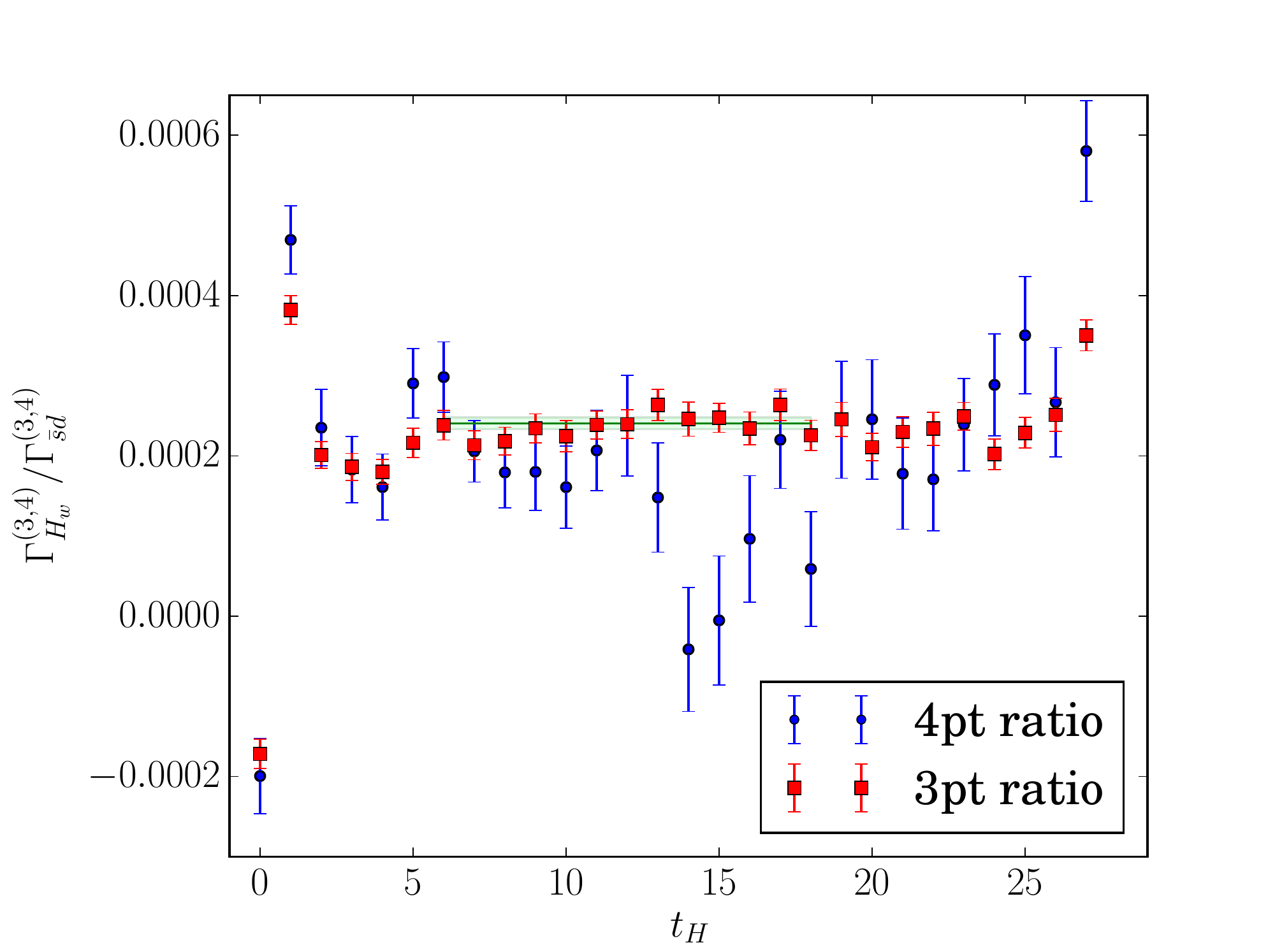}
	\end{center}
	\caption{\label{fig:c_s_fit}Determination of the parameter $c_s$ from a fit to the ratio of 3pt $H_W$ and $\bar{s}d$ correlators. The corresponding ratio of the 4pt correlator is also shown. The position of the plateau corresponds to $c_s=0.000240(8)$.}
\end{figure}

The first part of this analysis requires us to determine the parameter $c_s$. In Fig.~\ref{fig:c_s_fit} we show the determination of this parameter using Eq. (\ref{eq:c_s_ratio}) and either 3pt or 4pt functions. A cleaner signal is obtained from the ratio of 3pt functions, although the ratio of 4pt functions does also agree as expected for $t_K\ll t_H\ll t_J$ (albeit with much larger errors).

\begin{figure}
	\begin{center}
		\begin{tabular}{c c}
			\includegraphics[width=0.5\textwidth]{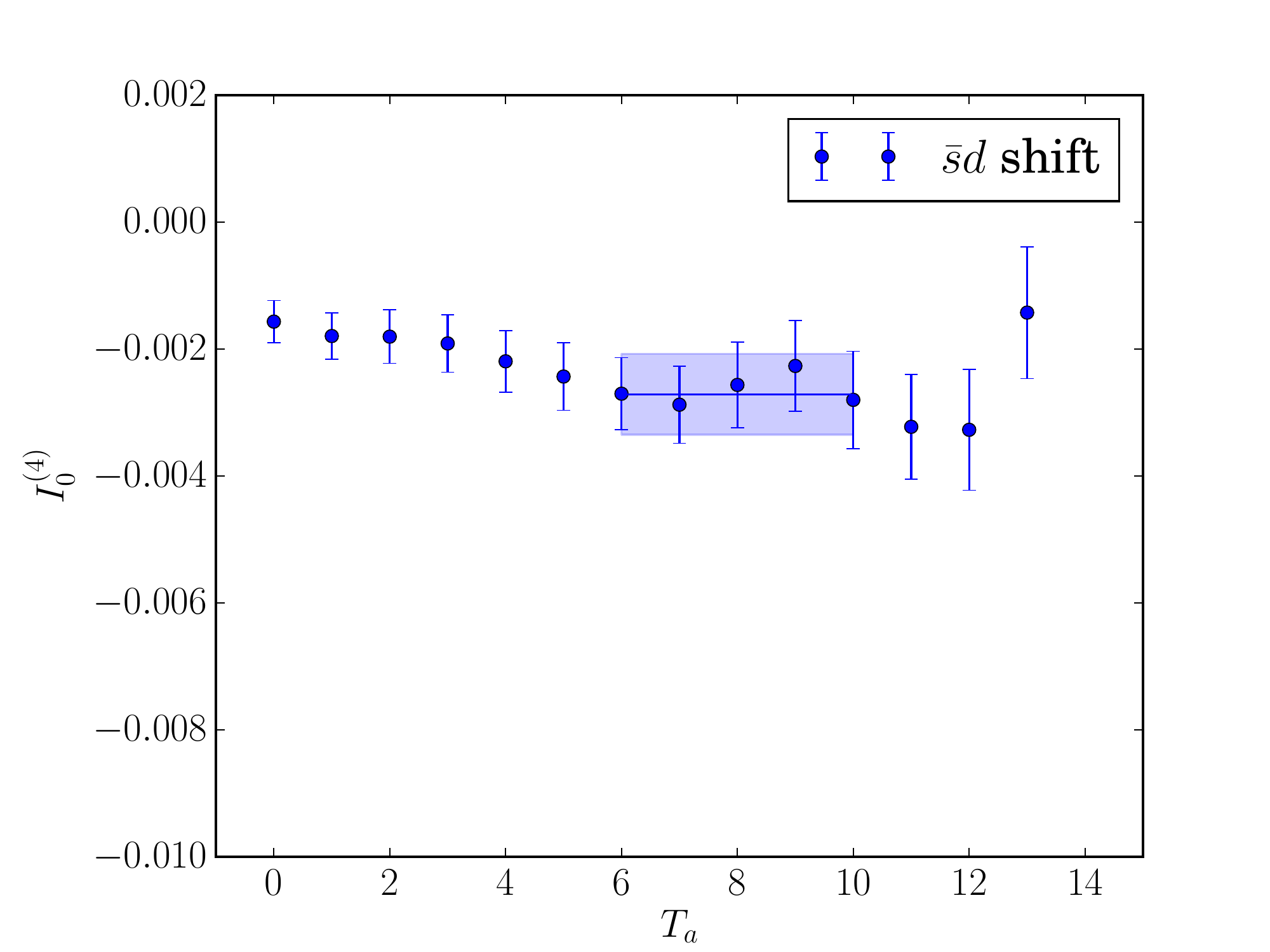} &
			\includegraphics[width=0.5\textwidth]{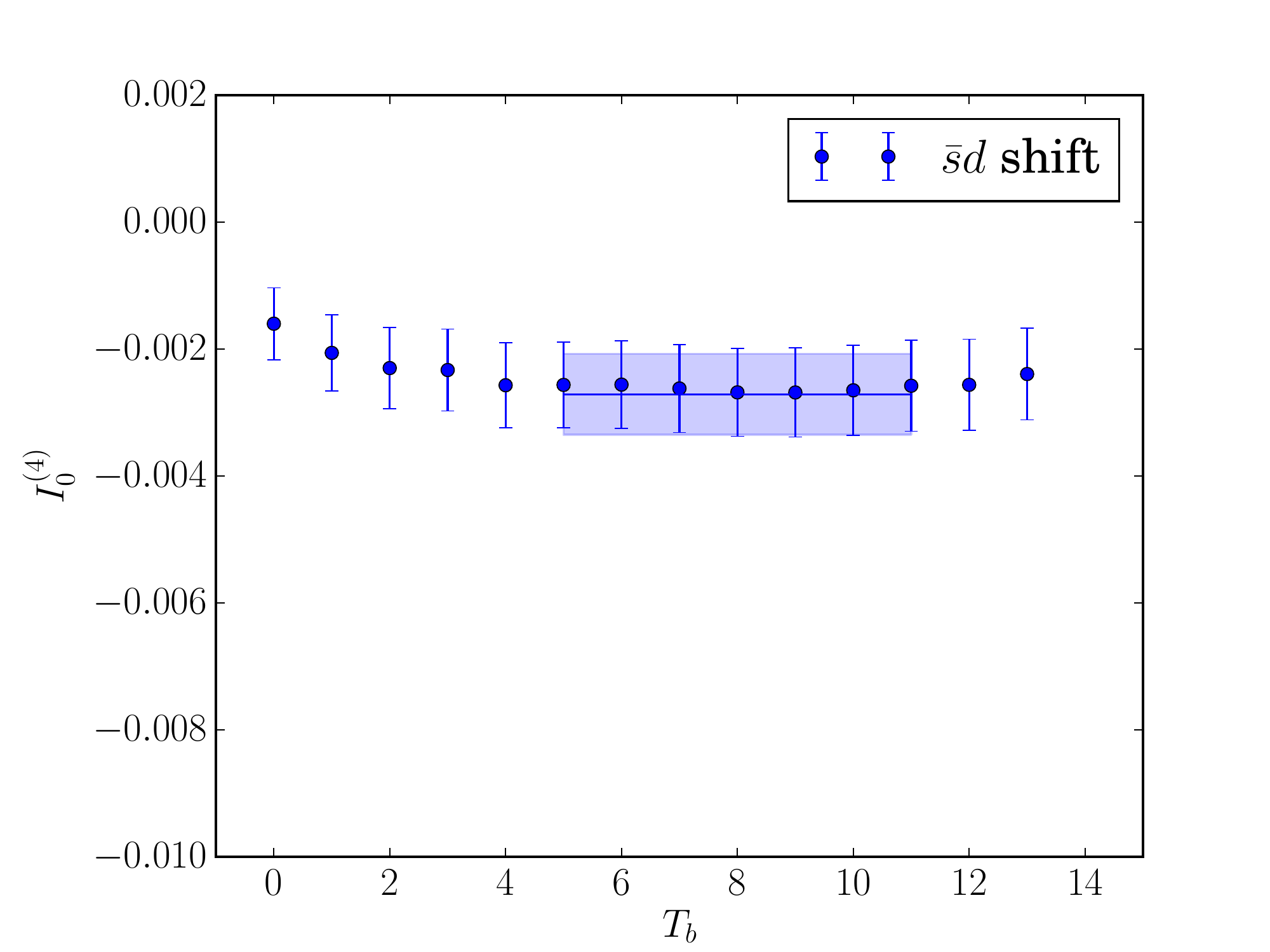} \\
			(a) & (b) \\
		\end{tabular}
	\end{center}
	\caption{\label{fig:method_2}The integrated 4pt correlator, shown for (a) $\int_{t_J-T_a}^{t_J+8}\tilde{\Gamma}^{(4)}_{0}-c_s\tilde{\Gamma}^{\bar{s}d(4)}_{0}dt_H$ to demonstrate the $T_a$ dependence and (b) $\int_{t_J-6}^{t_J+T_b}\tilde{\Gamma}^{(4)}_{0} - c_s\tilde{\Gamma}^{\bar{s}d(4)}_{0}dt_H$ to demonstrate the $T_b$ dependence. The single-pion exponential growth has been removed using method 2. The position of the plateau corresponds to $A_{0}=-0.0027(6)$, obtained by a fit to the data over the indicated range.}
\end{figure}

The resulting integrated correlator after shifting by the 4pt correlator with $H_W$ replaced by $\bar{s}d$ is shown in Fig.~\ref{fig:method_2}. We obtain the matrix element by fitting the correlator to a constant in the region where both sides of the integral plateau. We note that the $\bar{s}d$ shift appears to remove the decaying single kaon intermediate state contribution on the $T_b$ side of the integral, in addition to the single-pion exponential term. The reason for this appears to be that $c_{s}$ is very weakly dependent on the momentum, which can be understood from the fact that it is independent of momentum in the $SU(3)$ symmetric limit (cf. Appendix~\ref{sec:c_1_c_2_approx}).

\begin{figure}
	\begin{center}
		\begin{tabular}{c c}
			\includegraphics[width=0.5\textwidth]{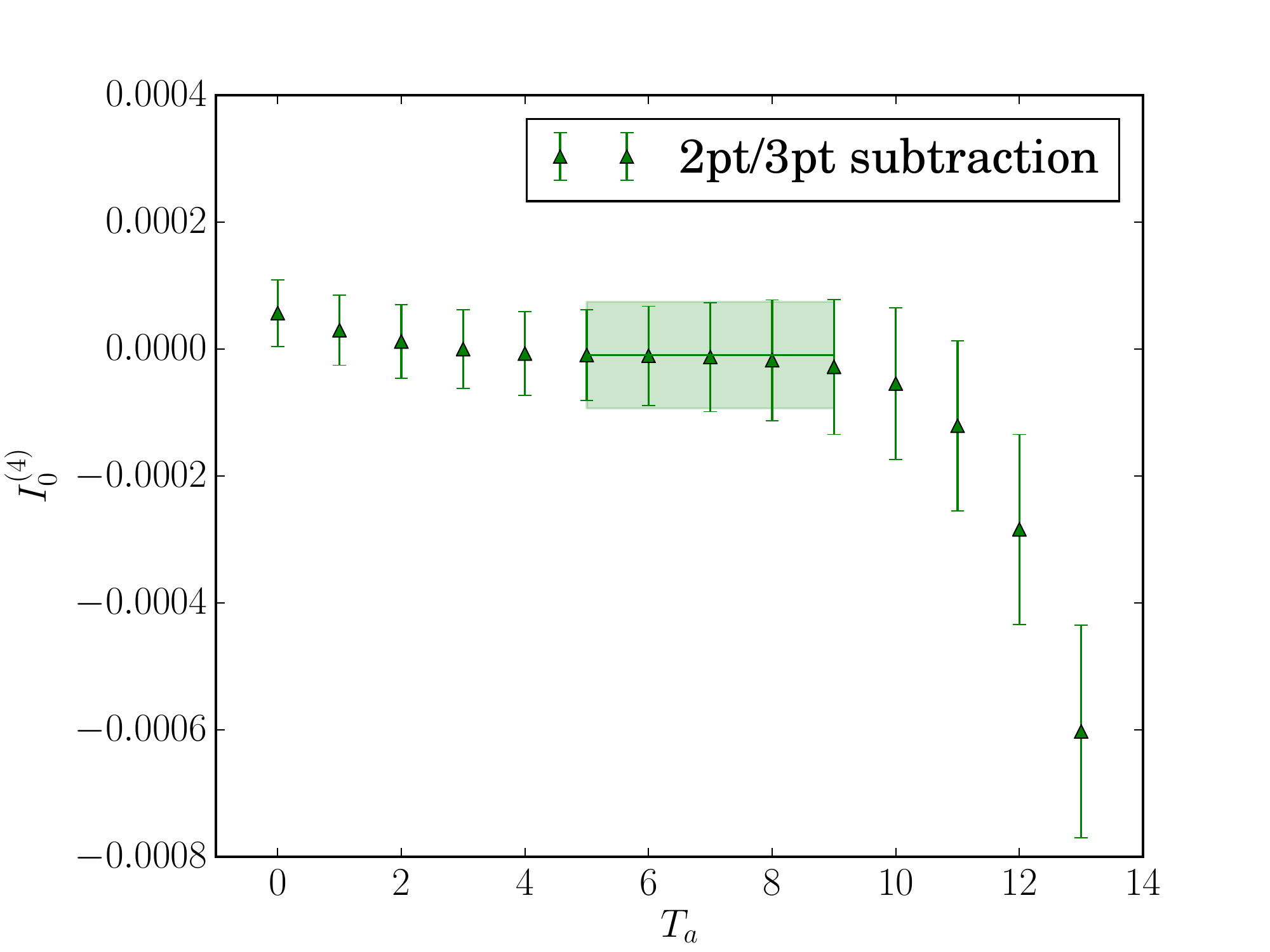} &
			\includegraphics[width=0.5\textwidth]{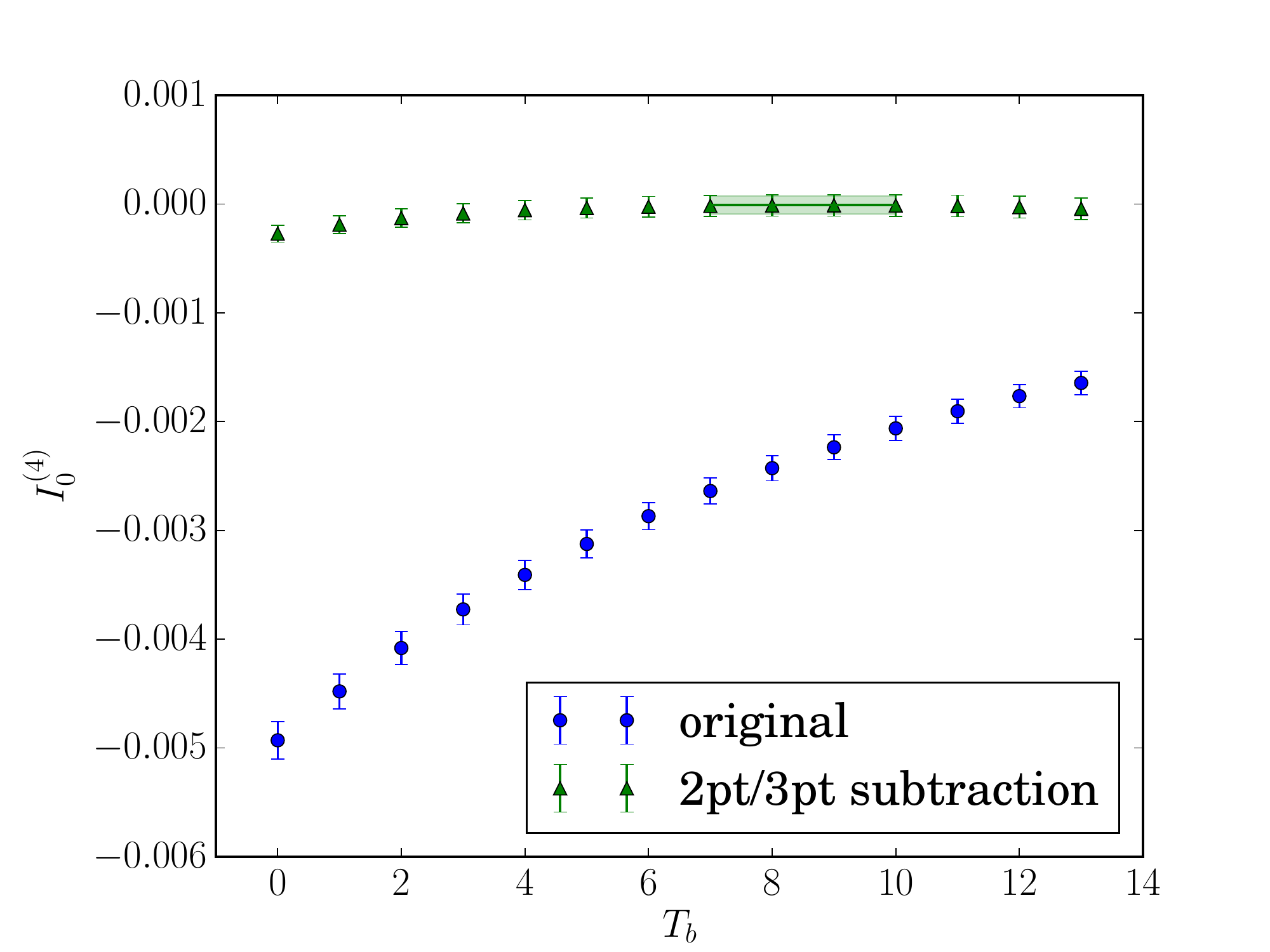} \\
			(a) & (b) \\
		\end{tabular}
	\end{center}
	\caption{\label{fig:method_1_for_sd}The integrated 4pt correlator with $H_W$ replaced by $\bar{s}d$, shown for (a) $\int_{t_J-T_A}^{t_J+8}c_s\tilde{\Gamma}^{\bar{s}d(4)}_{0}dt_H$ and (b) $\int_{t_J-6}^{t_J+T_B}c_s\tilde{\Gamma}^{\bar{s}d(4)}_{0}dt_H$. The single-pion exponential growth has been removed using method 1. The single kaon exponential decay has been removed using the approximation $\mathcal{M}_{\bar{s}d}\left(\mathbf{p}\right)=\mathcal{M}_{\bar{s}d}\left(\mathbf{k}\right)$. The position of the plateau corresponds to $A_{0}^{\bar{s}d}=-0.00001(8)$, obtained by a fit to the data over the indicated range.}
\end{figure}

\begin{figure}
	\begin{center}
		\includegraphics[width=0.65\textwidth]{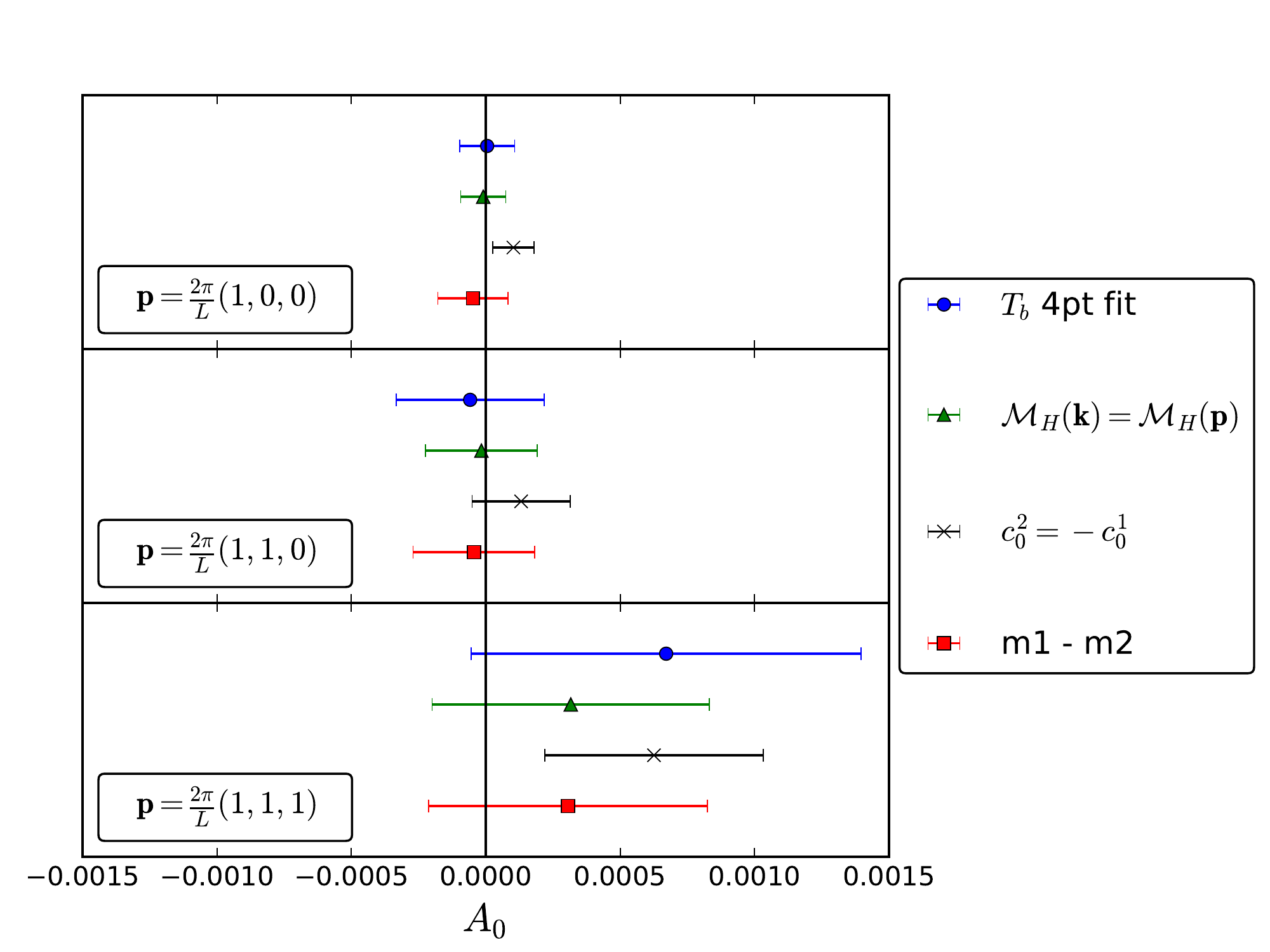}
	\end{center}
	\caption{\label{fig:sd_m1_results}Plot of the amplitudes (in lattice units) obtained using each of the different analysis methods for the $\Gamma^{(4)}_{\bar{s}d}$ correlator. The expected value of zero is marked explicitly.}
\end{figure}

An important test of this method is to check that the $\bar{s}d$ 4pt correlator gives no contribution to the final amplitude~\cite{Christ:2015aha}. As a consistency check, we can apply the 'method 1' integration techniques to this correlator in an attempt to verify that the matrix element contribution is consistent with zero. Plots of the integral of this correlator are shown in Fig.~\ref{fig:method_1_for_sd}, and the results for each pion momentum are displayed in Fig.~\ref{fig:sd_m1_results}. We remark that the result of these three analyzes are generally consistent with zero, as is the difference between the matrix elements obtained using either methods 1 or 2.

\subsection{Discussion}

A summary of the results of our analysis of the 4pt functions for the three choices of pion momenta studied is presented in Table~\ref{tab:results}. A comparison of statistical errors shows that both analysis methods 1 and 2 can be used to obtain the matrix element with similar statistical precision. The two methods also show remarkable agreement, suggesting that systematic effects are well controlled by our analysis. The two methods give the best agreement when we use the approximation $\mathcal{M}_H\left(\mathbf{k}\right)=\mathcal{M}_H\left(\mathbf{p}\right)$ in method 1 when constructing the coefficient $c_{0}^{2}$ of Eq. (\ref{eq:asymptotic_form}). This indicates that this approximation carries a smaller systematic error than the approximation $c_{0}^{1}=c_{0}^{2}$ for the choices of masses and pion momenta used in this simulation.

Our cleanest results are obtained when we used method 2 to perform the analysis, which does not use any approximations in the analysis process. Using these results we therefore compute the form factor for the decay using Eq. (\ref{eq:mat_elem_form_fac}). The results for the form factor are presented in Table~\ref{tab:form_factors}.

\begin{table}
	\begin{center}
	\begin{tabular}{|c|c|c|c|}
		\hline
		$\mathbf{p}$ & $\frac{2\pi}{L}\left(1,0,0\right)$ & $\frac{2\pi}{L}\left(1,1,0\right)$ & $\frac{2\pi}{L}\left(1,1,1\right)$ \tabularnewline
		\hline
		$z$ & $-0.5594(12)$ & $-1.0530(34)$ & $-1.4653(82)$ \tabularnewline
		$V(z)$ & $1.37(36)$ & $0.68(39)$ & $0.96(64)$ \tabularnewline
		\hline
	\end{tabular}
	\end{center}
	\caption{\label{tab:form_factors}The form factor of the $K\left(\mathbf{0}\right)\to\pi\left(\mathbf{p}\right)\gamma^{*}$ decay computed for the three pion momenta.}
\end{table}

It is instructive to perform our analysis separately for the loop diagrams $S$ and $E$, and the nonloop diagrams $W$ and $C$. While either combination of diagrams does not correspond to entire operators $Q_1$ and $Q_2$, it is useful to be able to study the diagrams involving the charm quark separately. The results are also shown in Table~\ref{tab:results}. We remark that we should find that $A_{\mu}=A_{\mu}^{C,W}+A_{\mu}^{S,E}$; it can be seen from the central values in Table~\ref{tab:results} that this generally holds well for all analysis methods. Small deviations from this relation represent a possible source of systematic error in our analysis procedure, which are introduced by using different choices of fit ranges for the individual diagrams rather than fitting the sum, and thus can be attributed to small excited state contaminations. Such errors however are significantly smaller than our statistical errors. An important observation to make is that even though the contribution of the single-pion intermediate state evidently contributes with opposite sign between the loop and nonloop diagrams (as seen in Fig.~\ref{fig:GIM_subtraction}), the four classes of diagrams all contribute constructively to the final matrix element. This is important from the perspective of our unphysical GIM cancellation: if we were to simulate with a heavier (thus more physical) charm quark we would expect the $S$ and $E$ diagrams to have a larger contribution and hence give us a more negative result for the matrix element. However we will leave a numerical test of the charm mass dependence until a future work, as this is not the primary focus of our present study.

Importantly, when simulations are performed with lighter values for $M_{\pi}$ and $M_K$, more states may contribute exponentially growing contributions (from $\pi\pi$ and $\pi\pi\pi$ intermediate states). It is instructive therefore to understand exactly how best to remove the single-pion state from simulations where it gives the only exponentially growing contribution. We have demonstrated the analysis techniques to remove this state cleanly with minimal systematic errors; hence it now remains to extend our simulations to physical masses such that the contributions of additional exponentially growing states can be investigated.

\section{\label{sec:Form_Factor}Form Factor}

\begin{figure}
	\begin{center}
		\includegraphics[width=0.4\linewidth]{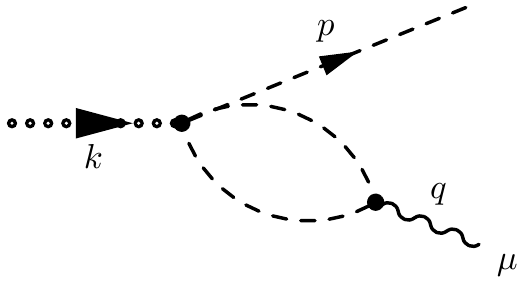}
	\end{center}
	\caption{\label{fig:3pi_rescatter}The one-loop contribution to the decays $K\to\pi\gamma^*$ arising as $\pi\pi\to\gamma^*$ rescattering in $K\to\pi\pi\pi$ decays.}
\end{figure}

One opportunity of lattice QCD is to test the previous work on rare kaon decays performed using effective theories such as $SU(3)$ ChPT. One previous analysis of the form factor~\cite{D'Ambrosio:1998yj} has led to a parametrization of the form
\begin{align}
V_{i}\left(z\right) = a_{i} + b_{i}z + V_{i}^{\pi\pi}\left(z\right), \label{eq:chipt_ansatz}
\end{align}
where $z=q^2/M_{K}^{2}$, and $V_{i}^{\pi\pi}\left(z\right)$ $\left(i=+,0\right)$ is introduced to account for $\pi\pi\to\gamma^*$ rescattering in $K\to\pi\pi\pi$ decays arising through the diagram show in Fig.~\ref{fig:3pi_rescatter}.  The most straightforward check is to test the relation Eq.\,(\ref{eq:chipt_ansatz}) by determining the constants $a_{i}$ and $b_{i}$ from simulation data. The contribution of the term $V_{i}^{\pi\pi}\left(z\right)$ is significantly smaller that the linear contribution for physical masses; for our initial calculation we can safely neglect this term. Experimentally the coefficients $a_{+}$ and $b_{+}$ have been determined from $K^+\rightarrow\pi^+\ell^+\ell^-$ spectra: $a_+=-0.578(16)$ and $b_+=-0.779(66)$ from $K^+\rightarrow\pi^+ e^+e^-$ data~\cite{Batley:2009aa} and $a_+=-0.575(39)$ and $b_+=-0.813(145)$ from $K^+\rightarrow\pi^+ \mu^+\mu^-$ data~\cite{Batley:2011zz}. 

The parametrization of Eq. (\ref{eq:chipt_ansatz}) is expected to be a good approximation to the $\mathcal{O}\left(p^6\right)$ ChPT form factor. It is already well known that existing $\mathcal{O}\left(p^4\right)$ ChPT predictions~\cite{Ecker:1987qi} for the parameter $b_+$ do not correctly predict experimental observations~\cite{D'Ambrosio:1998yj, Friot:2004yr}. Analysis of this decay in ChPT up to $\mathcal{O}\left(p^4\right)$ gives the following predictions for the coefficients $a_{i}$ and $b_{i}$~\cite{D'Ambrosio:1998yj},
\begin{eqnarray}
a_{+} = \dfrac{G_8}{G_F}\left(\dfrac{1}{3} - w_+ \right),\label{eq:a_mass_dep} & \:\: &
a_{0} = -\dfrac{G_8}{G_F}\left(\dfrac{1}{3} - w_0 \right), \\
b_{+} = -\dfrac{G_8}{G_F}\dfrac{1}{60}, & \:\: &
b_{0} = \dfrac{G_8}{G_F}\dfrac{1}{60},
\end{eqnarray}
where $w_{i}$ are defined in terms of $SU(3)$ low energy constants (LECs) $N_{14}^{r}(\mu)$, $N_{15}^{r}(\mu)$ and $L_{9}^{r}$ as
\begin{eqnarray}
w_+ &=& \dfrac{64\pi^2}{3}\left(N_{14}^{r}(\mu) - N_{15}^{r}(\mu) + 3L_{9}^{r}(\mu)\right) + \dfrac{1}{3}\ln\left(\dfrac{\mu^2}{M_{K}M_{\pi}}\right), \label{eq:a_ChPT_parms}\\
w_0 &=& \dfrac{32\pi^2}{3}\left(N_{14}^{r}(\mu) + N_{15}^{r}(\mu)\right) + \dfrac{1}{3}\ln\left(\dfrac{\mu^2}{M_{K}^{2}}\right)
\label{eq:ChPT_parms}
\end{eqnarray}
for some renormalization scale $\mu$. The coefficient $b_{+}$ depends only on the LEC $G_8$, which can be determined using information from $K\to\pi\pi$ decay amplitudes~\cite{Cirigliano:2004a}. A comparison with the experimental result thus demonstrates that large corrections must be expected at $\mathcal{O}\left(p^6\right)$. Models that go beyond $\mathcal{O}\left(p^4\right)$ ChPT in an attempt to make predictions for $b_{+}$ have been proposed~\cite{Friot:2004yr,Dubnickova:2006mk}, although such models depend heavily on vector meson masses and thus a comparison with our lattice data is difficult. 

\begin{figure}
	\begin{center}
		\includegraphics[scale=0.5]{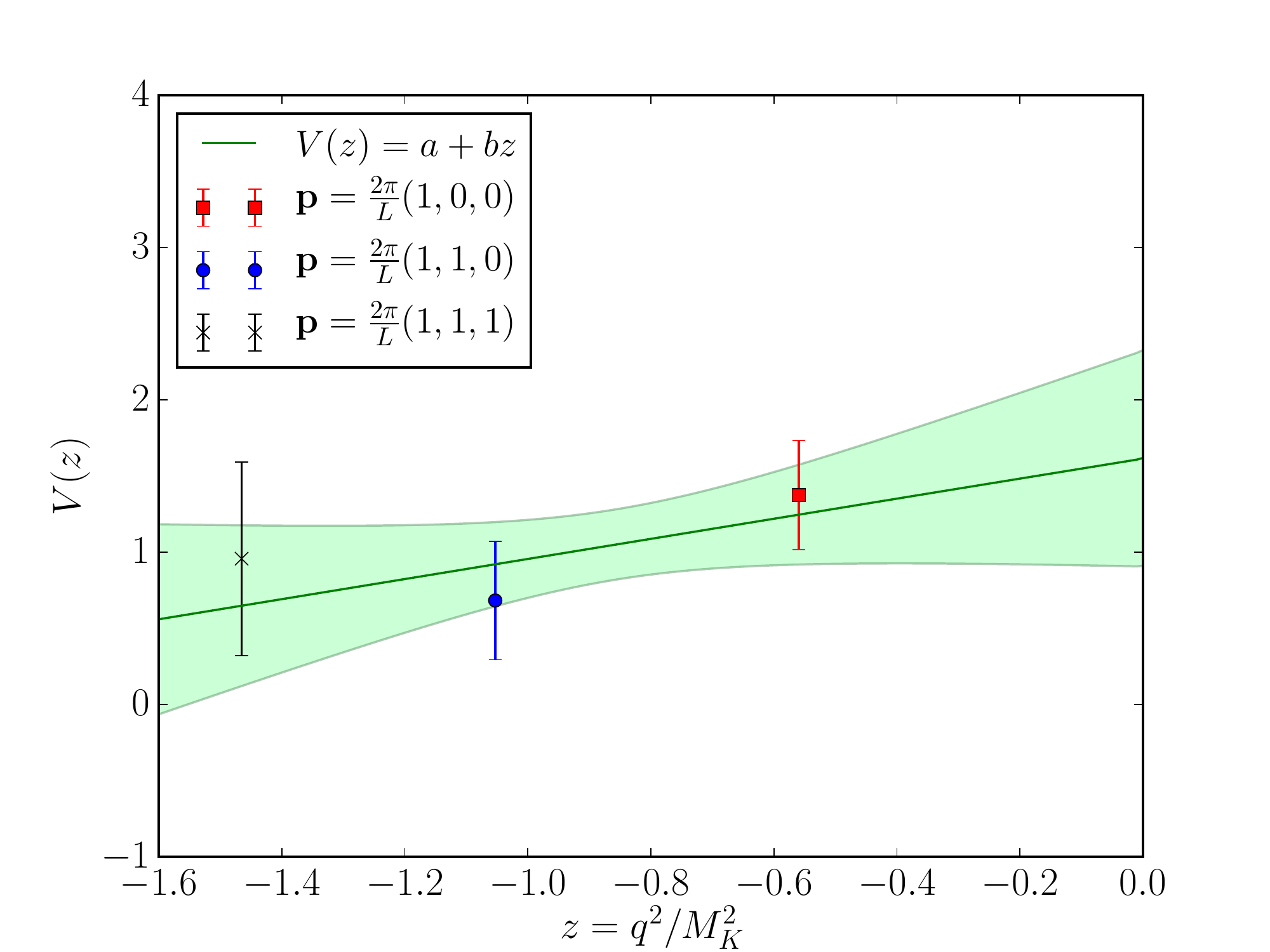}
	\end{center}
	\caption{\label{fig:form_factor}Dependence of the form factor for the decay $K^+\rightarrow\pi^+\ell^+\ell^-$ upon $z=q^2/M_{K}^{2}$. Our lattice data is fit to a linear ansatz to obtain $a=1.6(7)$ and $b=0.7(8)$.}
\end{figure}

In Fig.~\ref{fig:form_factor} we display the dependence of the form factor extracted from lattice data upon $z=q^2/M_{K}^{2}$. Although our simulation takes place with highly unphysical masses of the pion and kaon, we are able to make some insights. Since we have only three data points at quite large spacelike momenta, we will not be able to fully explore the ChPT anastz in Eq.~(\ref{eq:chipt_ansatz}). Here we simply use a linear fit, which does provide a reasonable description of our data with a $\chi^2/\mathrm{d.o.f}=0.74$. The parameters we obtain, $a^{lat}_{+}=1.6(7)$ and $b^{lat}_{+}=0.7(8)$, are different from the parameters obtained from phenomenological fits to experimental data, $a_{+}^{exp}=-0.578(16)$ and $b_{+}^{exp}=-0.779(66)$. However such a comparison must be taken with care given the unphysical masses used in our simulation.

The most relevant and interesting comparison we make with experimental results at this stage is to note that the sizes of the absolute errors on the parameters $a_{+}$ and $b_{+}$ obtained via our lattice calculation are at least an order of magnitude larger than those obtained from fits to experimental data. As an exploratory study our aim has been to evaluate the feasibility of this calculation, which we have done successfully. In the future we foresee a greater expenditure of computer time in order to produce significantly smaller errors in a calculation closer to the physical point.

\section{\label{sec:Conclusion}Conclusions}

In this paper we have demonstrated that it is possible to calculate the long-distance contributions to $K^+\to\pi^+\ell^+\ell^-$ amplitudes arising from the connected diagrams using lattice QCD.  The connected diagrams are expected to provide the dominant contribution.  However we expect that with a substantially increased statistical sample, the methods developed here will also be able to determine the disconnected part.  The extraction of these amplitudes is made difficult by the presence of intermediate states that are lighter than the kaon, leading to unphysical, exponentially growing contributions. We have employed two different methods for removing these unphysical parts, which grow exponentially as the range for the integral over the time separation between the electromagnetic current and the effective weak Hamiltonian is increased.  Both methods successfully remove this unwanted contribution and give consistent results for the physical amplitude.  The stage is now set for a calculation with lighter kaon and pion masses, in particular chosen such that the $\pi\pi$ states will also contribute exponentially growing terms. This will allow us to develop our analysis techniques further, in particular those needed to handle the contribution of these additional exponentially growing terms, and to make comparisons to experimental results. 

We emphasise that our analysis techniques are also applicable to the neutral decay $K_S\to\pi^0\ell^+\ell^-$.  While we have been unable to resolve a signal for this amplitude in our present calculation because of the additional disconnected contractions needed for this decay, we expect that with a larger sample of configurations and additional variance reduction methods, the matrix elements for this decay should be accessible with a precision similar to what was obtained here for the $K^+\to\pi^+\ell^+\ell^-$  amplitudes.

Although our calculation has been performed with unphysical values for the kaon and pion masses, it is nevertheless interesting to make qualitative comparisons to the expectations from chiral perturbation theory. As a schematic calculation, we have tested using $\mathcal{O}(p^4)$ ChPT formulas to extrapolate our results to physical pion and kaon masses to compare with experimental data.  For both the lattice and experimental results a ChPT-motivated fit ansatz can be used to produce values of $V_+(0)$, which is known at $\mathcal{O}(p^4)$. Our lattice result [$a^{lat}_+=1.6(7)$] cannot reasonably be compared to the experimental result [$a^{exp}_+ = -0.578(16)$] at this stage, as our simulations use meson masses that are considerably heavier than their physical values. As we begin to simulate with lighter pion and kaon masses, we will be able to study the mass dependence of $a_+$ and $b_+$ and ultimately at the physical point we can make direct comparisons with experiment.  It is important to note that the size of the errors on the results of our calculation are an order of magnitude greater than those obtained from fits to experimental results.  However, we expect that significant reductions in our statistical errors will be possible by increasing the number of gauge samples that are studied, expanding the number of calculations performed on each sample and employing further variance reduction methods such as all-mode averaging~\cite{Bali:2009hu, Blum:2012uh}  and all-to-all propagators~\cite{Foley:2005ac}. 

As mentioned above, we plan to extend this calculation to lighter and ultimately physical up and down quark masses.  However, a significant barrier which must be overcome in a truly physical calculation is a proper treatment of the charm quark.  Our 533 MeV choice for the charm quark mass provides a conservative environment to explore computational methods and determine statistical uncertainties.  However, using the physical $1.3$ GeV value for $m_c$ poses substantial computational costs since we must use both a sufficiently small lattice spacing to properly treat this large mass and a sufficiently large volume to properly treat a physical pion mass.  This difficulty associated with the large difference in the energy scales of charm and light quarks could be avoided if we choose to integrate out the charm quark and work in the three-flavor theory.  In this approach the GIM cancellation treated nonperturbatively here would be dealt with using QCD perturbation theory, resulting in an expanded set of four-quark effective operators including new gluonic and photonic penguin operators whose coefficient would be determined in perturbation theory.  This treatment would be very similar to recent, three-flavor calculations of $K\to\pi\pi$ decay~\cite{Blum:2015ywa, Bai:2015nea}.   However, the results of Inami and Lim~\cite{ Inami:1980fz} for the case where the electromagnetic vertex is inserted in the GIM-subtracted quark loop in the $S$ and $E$ graphs of Fig.~\ref{fig:H_W_contractions} suggest that such charm quark contributions may be a substantial fraction of the $K^+\to\pi^+\ell^+\ell^-$ decay amplitude.  As a result, we anticipate that a nonperturbative treatment of the charm quark may be necessary as is the case for a similar charm quark contribution to the $K_L-K_S$ mass difference.  Thus, a physical calculation of both the $K^+\to\pi^+\ell^+\ell^-$ and $K_s\to\pi^0\ell^+\ell^-$ decays should become possible in the next three to four years when the next generation of computers becomes available.

\section*{Acknowledgements}

We gratefully acknowledge many helpful discussions with our colleagues from the RBC and UKQCD Collaborations. N.H.C. and X.F. are supported by US DOE Grant No. \#DE-SC0011941. A.J. acknowledges the European Research Council under the European Community's Seventh Framework Programme (FP7/2007-2013) ERC Grant agreement No. 279757. A.L. is supported by an EPSRC Doctoral Training Centre Grant No. EP/G03690X/1. A.P. and C.T.S. are supported by UK STFC Grant No. ST/L000296/1 and A.P. additionally by Grant No. ST/L000458/1.


\appendix

\section*{Appendix}

\section{Calculation of Propagators\label{sec:props}}

In this section we give detailed explanations of the computation of the propagators used in our simulation, being specific where necessary to the case of domain wall fermions.

\subsection{Random volume source propagator\label{sec:random_vol}}

The propagator we use in our calculation to compute quark loops is $S(x,x)^{\alpha,\alpha}_{a,a}$, where the roman index $a$ denotes color indices and the greek index $\alpha$ denotes spin indices. It is defined via
\begin{align}
S(x,x)^{\alpha,\alpha}_{a,a}=&D^{-1}(x,x)^{\alpha,\alpha}_{a,a} \nonumber
\\ = & \sum_{y,\beta,b} D^{-1} (x,y)^{\alpha,\beta}_{a,b} \delta(y-x)\delta^{\alpha,\beta}\delta_{a,b}\nonumber\\
= & \sum_{y,\beta,b}\left\langle D^{-1}(x,y)^{\alpha,\beta}_{a,b}\eta(y)\eta^{*}(x)\delta^{\alpha,\beta}\delta_{a,b}\right\rangle _{\eta},\label{eq:noise_sources}
\end{align}
where $D$ is the Dirac operator and $\eta(x)$ are vectors of random complex numbers that satisfy the constraints~\cite{Bernardson:1993he}
\begin{align}
\left|\eta(x)\right|^{2} = 1,\:
\left\langle \eta(x)\right\rangle _{\eta} = 0,\:
\left\langle \eta(y)\eta^{*}(x)\right\rangle _{\eta} = \delta(y-x).
\end{align}
We have used the notation $\left\langle \cdots\right\rangle _{\eta}$ to indicate the stochastic average over a large number of noise vectors to distinguish it from the usual gauge average. To satisfy Eq.~(\ref{eq:noise_sources}) we take $\eta(x)$ to be constant across all spin and color indices corresponding to a single site. We use complex $\mathbb{Z}_{2}$ noise to generate the vectors $\eta(x)$~\cite{Dong:1993pk,Foster:1998vw}.

\subsection{Sequential propagator\label{sec:seq_prop}}

In this section we introduce the calculation of a sequential propagator for a Shamir domain wall fermion~\cite{Shamir:1993yf,Furman:1994ky}. While the physical fermion fields exist in four-dimensional spacetime, the conserved domain wall current we must consider exists in five-dimensional spacetime. Let us first define the five-dimensional fermion fields, $\Phi\left(s,x\right)$, where $s$ indexes the position in the fifth dimension, $s=1,...,L_{s}$. We define the "physical," four-dimensional quark-fields, $\psi\left(x\right)$, as chiral projections of the five-dimensional fields $\Phi$, i.e.
\begin{align}
\psi\left(x\right)=P_{R}\Phi\left(x,L_s\right)+P_{L}\Phi\left(x,1\right),
\end{align}
where $P_R$ and $P_L$ are the right and left projection operators respectively. 

Before we discuss the sequential propagator we first introduce propagators from the surface field $\psi$ into the five-dimensional bulk and vice versa, i.e.
\begin{align}
S_{SB}\left(x,s;y\right)&=\left\langle \Phi\left(x,s\right) \overline{\psi}\left(y\right) \right\rangle, \\
S_{BS}\left(x;y,s\right)&=\left\langle \psi\left(x\right) \overline{\Phi}\left(y,s\right) \right\rangle.
\end{align}
The five-dimensional conserved current for the domain wall action is made up of the following components~\cite{Shamir:1993yf,Furman:1994ky}: for the first four dimensions we have
\begin{align}\label{eq:cons_mu1-4}
j_{\mu}\left(x,s\right)=\dfrac{1}{2}\left(\overline{\Phi}\left(x,s\right)\left(1+\gamma_{\mu}\right)U_{\mu}\left(x\right)\Phi\left(x+\hat{\mu},s\right)-\overline{\Phi}\left(x+\hat{\mu},s\right)\left(1-\gamma_{\mu}\right)U_{\mu}^{\dagger}\left(x\right)\Phi\left(x,s\right)\right)
\end{align}
where $U_{\mu}\left(x\right)$ is the link variable in the direction $\mu$, and $\hat{\mu}$ is the unit vector in the direction $\mu$. For completeness the fifth component is given by
\begin{align}\label{eq:cons_mu5}
j_{5}\left(x,s\right)=\overline{\Phi}\left(x,s\right)P_{R}\Phi\left(x,s+1\right)-\overline{\Phi}\left(x,s+1\right)P_{L}\Phi\left(x,s\right),
\end{align}
although it is unnecessary for our calculation. The overall four-dimensional conserved current is given by the expression
\begin{align}\label{eq:4D_cons_curr}
J_{\mu}\left(x\right)=\sum_{s}j_{\mu}\left(x,s\right). 
\end{align}

Putting this together, we must calculate
\begin{align}
\Sigma_{\mu}\left(x,x_{0};y,y_{0}\right) = \sum_{\mathbf{z},s}e^{i\mathbf{p}\cdot\mathbf{z}}S_{BS}(x_0,\mathbf{x};t_{J},\mathbf{z},s) K_{\mu}\left(t_{J},\mathbf{z},s\right) S_{SB}(t_{J},\mathbf{z},s;y_0,\mathbf{y}),
\end{align}
where $K_{\mu}$ is the kernel of the conserved current that follows from Eqs. (\ref{eq:cons_mu1-4})-(\ref{eq:4D_cons_curr}). This propagator is obtained from an additional inversion by solving
\begin{align}
\sum_{x}D\left(t_J,\mathbf{z},s;x_0,\mathbf{x}\right)\Sigma_{\mu}\left(x_0,\mathbf{x},;y_0,\mathbf{y}\right)= e^{i\mathbf{p}\cdot\mathbf{z}}K_{\mu}\left(t_{J},\mathbf{z},s\right)S_{SB}(t_{J},\mathbf{z},s;y_0,\mathbf{y}),
\end{align}
for $\Sigma_{\mu}$ where $D$ is the five-dimensional Dirac operator.

Lastly we comment on the $\gamma_5$ Hermiticity properties of this propagator. In general we find that for a sequential propagator with an operator insertion $\mathcal{O}$, we have
\begin{align}
\Sigma_{\mathcal{O}}\left(x,y\right)=\gamma_5 \Sigma_{\mathcal{O}^{\dagger}}^{\dagger}\left(x,y\right)\gamma_5.
\end{align}
For the example of the vector current, we simply have
\begin{align}
\Sigma_{\mu}\left(x,y\right)=-\gamma_5 \Sigma_{\mu}^{\dagger}\left(x,y\right)\gamma_5.
\end{align}

\section{\label{sec:c_1_c_2_approx}Approximations}

\subsection{$c^{2}_{0}(\mathbf{k})=-c^{1}_{0}(\mathbf{k})$}

In this section we provide a justification for the approximation $c^{2}_{0}\left(\mathbf{k},\mathbf{p}\right) =-c^{1}_{0}\left(\mathbf{k},\mathbf{p}\right)$. The basis of this approximation is the identification that the relation holds exactly when $\mathbf{k}=\mathbf{p}$. To show this, we define
$c^{1}_{0}(\mathbf{k})$ and $c^{2}_{0}(\mathbf{k})$ respectively as
\begin{equation}
c_{0}^{1}\left(\mathbf{k}\right)=\dfrac{\mathcal{M}_{0}^{J,\pi}\left(\mathbf{0}\right)\mathcal{M}_{H}\left(\mathbf{k}\right)}{2E_{\pi}\left(\mathbf{k}\right)\left(E_K\left(\mathbf{k}\right)-E_{\pi}\left(\mathbf{k}\right)\right)},\:
c_{0}^{2}\left(\mathbf{k}\right)=-\dfrac{\mathcal{M}_{0}^{J,K}\left(\mathbf{0}\right)\mathcal{M}_{H}\left(\mathbf{k}\right)}{2E_{K}\left(\mathbf{k}\right)\left(E_K\left(\mathbf{k}\right)-E_{\pi}\left(\mathbf{k}\right)\right)}.
\label{eq:c_1_c_2_app}
\end{equation}
In general the current matrix element can be decomposed as
\begin{align}
\mathcal{M}_{\mu}^{J,P}\left(k,p\right)=\left(k+p\right)_{\mu}F^{P}\left((k-p)^2\right),
\end{align}
where $F^{P}$ is the electromagnetic form factor of the particle $P$. At the point $k=p$, we find that
\begin{align}
\mathcal{M}_{0}^{J,P}\left(k,k\right)=2E_{P}\left(\mathbf{k}\right)Q,
\end{align}
where $E_{P}$ is the energy of the particle in question, and $Q$ is its charge (in units of the elementary charge). The factor of $2E_{P}\left(\mathbf{k}\right)$ is canceled by the normalization factor in both $c^{1}_{0}(\mathbf{k})$ and $c^{2}_{0}(\mathbf{k})$. The remaining factor of
\begin{align}
\dfrac{\mathcal{M}_{H}\left(\mathbf{k}\right)}{E_K\left(\mathbf{k}\right)-E_{\pi}\left(\mathbf{k}\right)}
\end{align}
is common to both $c^{1}_{0}(\mathbf{k})$ and $c^{2}_{0}(\mathbf{k})$; it is thus clear to see that $c^{2}_{0}(\mathbf{k})=-c^{1}_{0}(\mathbf{k})$.

\subsection{$SU(3)$ symmetric limit}

In this section we show that the quantity $\left\langle \pi\left(\mathbf{p}\right) \right|H_{W}\left|K\left(\mathbf{p}\right)\right\rangle$ is independent of momentum in the $SU(3)$ symmetric limit. Let us consider the matrix element
\begin{align}
\mathcal{M}_{H}\left(\mathbf{p}\right)&=\left\langle \pi\left(\mathbf{p}\right) \right|H_{W}\left|K\left(\mathbf{p}\right)\right\rangle \\
&= \left\langle \pi\left(\mathbf{k}\right) \right|B^{-1}_{\pi}\left(k,p\right) H_{W} B_{K}\left(k,p\right) \left|K\left(\mathbf{k}\right)\right\rangle
\end{align}
where $B_{P}\left(k,p\right)$ is the boost into the frame where the particle $P$ has momentum $\mathbf{k}$ from the frame where it has momentum $\mathbf{p}$. The Lorentz boost depends on the particle's mass, and so in general the quantity
\begin{align}
B^{-1}_{\pi}\left(k,p\right) H_{W} B_{K}\left(k,p\right)
\end{align}
cannot be trivially decomposed. However when we take the limit $M_K\to M_{\pi}$, we find that
\begin{align}
B^{-1}_{\pi}\left(k,p\right) H_{W} B_{\pi}\left(k,p\right)=H_{W}.
\end{align}
The equality holds because the operator $H_{W}$ is a Lorentz scalar. It holds therefore that in the $SU\left(3\right)$ symmetric limit, the matrix element $\left\langle \pi\left(\mathbf{p}\right) \right|H_{W}\left|K\left(\mathbf{p}\right)\right\rangle$ is independent of momentum. A similar argument is true also for $\left\langle \pi\left(\mathbf{p}\right) \right|\bar{s}d\left|K\left(\mathbf{p}\right)\right\rangle$. As a result the ratio $c_s$ [Eq. (\ref{eq:c_s_ratio})] is also independent of momentum in this limit.

\bibliography{rare_kaon_numerical_paper}
\bibliographystyle{JHEP2}

\end{document}